\documentclass[a4,letterpaper,11pt]{article}

\pdfoutput=1 

\usepackage{jheppub} 
\usepackage[T1]{fontenc}
\usepackage{lmodern}
\usepackage{amsmath}
\usepackage{amssymb, bm}
\usepackage{graphicx}
\usepackage{xspace}
\usepackage{multirow,footnote}
\usepackage{booktabs}
\usepackage[absolute]{textpos}

\newcommand{\GeV}{\text{ GeV}}
\newcommand{\as}{\alpha_s}
\newcommand{\MSbar}{\overline{\text{MS}}}
\newcommand{\msm}{\overline{m}}

\newcommand{\gam}{\gamma}

\newcommand{\cO}{\mathcal{O}}
\newcommand{\cM}{\mathcal{M}}
\newcommand{\cP}{\mathcal{P}}

\newcommand{\nn}{\nonumber} 

\newcommand{\be}{\begin{equation}}
\newcommand{\ee}{\end{equation}}
\newcommand{\bea}{\begin{eqnarray}}
\newcommand{\eea}{\end{eqnarray}}
\newcommand{\bse}{\begin{subequations}}
\newcommand{\ese}{\end{subequations}}

\newcommand{\eq}[1]{Eq.~\eqref{#1}}
\newcommand{\eqs}[2]{Eqs.~\eqref{#1} and \eqref{#2}}
\newcommand{\eqss}[3]{Eqs.~\eqref{#1}, \eqref{#2}, and \eqref{#3}}

\renewcommand{\sec}[1]{Sec.~\ref{sec:#1}}

\newcommand{\appx}[1]{App.~\ref{app:#1}}\newcommand{\fig}[1]{Fig.~\ref{#1}}

\newcommand{\tab}[1]{Table~\ref{#1}}

\definecolor{orange}{rgb}{1,0.5,0}

\newcommand{\sing}{\text{sing}}
\newcommand{\ns}{\text{ns}}

\begin{document}

\title{ \Large Pseudoscalar Quarkonium$\bm{+\gamma}$ Production at NLL+NLO accuracy}
\author[a]{Hee Sok Chung,}
\author[b]{June-Haak Ee,}
\author[c]{Daekyoung Kang,}
\author[b]{U-Rae Kim,}
\author[b]{Jungil Lee,}
\author[c]{Xiang-Peng Wang}

\affiliation[a]{Physik Department, Technische Universit\"at M\"unchen, 
\\James-Franck-Str. 1, 85748 Garching, Germany}
\affiliation[b]{Department of Physics, Korea University, 
\\Seoul 02841, Korea}
\affiliation[c]{Key Laboratory of Nuclear Physics and Ion-beam Application (MOE) and Institute of Modern Physics, Fudan University, 
\\Shanghai 200433, China}

\emailAdd{heesok.chung@tum.de}
\emailAdd{chodigi@gmail.com}
\emailAdd{dkang@fudan.edu.cn}
\emailAdd{sadafada@korea.ac.kr}
\emailAdd{jungil@korea.ac.kr}
\emailAdd{xiang.peng.wang@desy.de}

\abstract{ 
We consider the exclusive pseudoscalar heavy-quarkonium ($\eta_{b,c}$) production in association with a photon at  future lepton colliders where the collider energies of $O(10^2)$\,GeV are far greater than the quarkonium mass.
At these energies, the logarithm of mass to collision energy becomes increasingly large hence its resummation becomes particularly important. By making use of the light-cone-distribution factorization formula, we resum the logarithms up to next-to-leading-logarithmic accuracy (NLL) that corresponds to order-$\as$ accuracy. We combine the resummed result with a known fixed-order result at next-to-leading order (NLO)  such that both resummed-logarithmic terms and non-logarithmic terms are included at the same order  in $\as$. This allowed us to provide reliable predictions at accuracies of order $\alpha_s$ ranging from relatively low energies near quarkonium mass to the collider energies of $O(10^2)$\,GeV.
We also include the leading relativistic corrections resummed at leading-logarithmic accuracy. 
Our prediction at the Belle energy is comparable with fixed-order predictions in literatures while it shows a large deviation from a recent Belle's upper limit by about $4\,\sigma$. 
Finally, we make predictions for the energies of future $Z$- and Higgs factories.}

\maketitle
\section{Introduction}
\label{sec:Intro}
Rigorous quantitative understanding of the heavy-quarkonium production at high-energy colliders \cite{Bodwin:2013nua} is a key probe not only to the features of quantum chromodynamics (QCD) but also to fundamental phenomena such as quark-gluon plasma (QGP) in heavy-ion collisions \cite{Andronic:2015wma} and heavy-quark Yukawa couplings to Higgs \cite{Sirunyan:2018fmm,Aaboud:2018txb}. An effective field-theoretic framework called the nonrelativistic QCD (NRQCD) \cite{Bodwin:1994jh} can be employed to predict quarkonium productions at high-energy colliders in a systematic way. NRQCD describes the dynamics inside a quarkonium at the energy scale $m_Q v^2$, where $m_Q$ is 
the mass of the heavy quark $Q$  and $v$ is the relative velocity of the $Q$ and $\bar{Q}$ in the bound state. 
NRQCD is blind to the short-distance dynamics at higher energy scales of order $\gtrsim m_Q$ and the corresponding short-distance coefficients
 can be determined by matching to the full theory, QCD, which is known to be correct in all accessible energy scales. 
As a result, the production cross sections or decay rates involving heavy quarkonia can be expressed as linear combinations of NRQCD long-distance matrix elements (LDME) with the short-distance coefficients.

Exclusive processes such as associated photon production  or double-quarkonium production at the lepton colliders like $B$ factories and BES have been extensively studied in the framework of NRQCD. Future lepton colliders such as ILC\cite{Behnke:2013xla}, CEPC\cite{CEPCStudyGroup:2018ghi}, and FCC-ee\cite{Gomez-Ceballos:2013zzn} offer opportunities to test our understanding of their productions at higher energies of $O(10^2)$\,GeV.
In a collision at such a large center-of-momentum (CM) energy $\sqrt{s}$, the cross section of a quarkonium has an uncomfortably
strong dependence on the large logarithm of the ratio 
\be
r\equiv \frac{4m_Q^2}{s}.
\ee
A straightforward extrapolation of the prediction for a lower-energy process  of $\sqrt{s}\lesssim 10$\,GeV to  higher-energy processes listed above may result in a failure of predictive power. Thus the accuracy of a prediction can be reasonably controlled only after resumming the large logarithmic contributions in a proper way because such a logarithm cannot be suppressed by the strong coupling constant:
\be\label{eq:asLog}
\alpha_s \ln r \sim O(1).\nn
\ee
The resummation of such a logarithm can be made by employing the light-cone (LC) approach  \cite{Lepage:1980fj,Chernyak:1983ej} or, equivalently, the soft-collinear effective theory (SCET) \cite{Bauer:2002nz}. In SCET, the scattering amplitude or current-current correlator is factorized into the following factors: the hard-scattering kernel involving scales of $\sqrt{s}$, the light-cone distribution amplitude (LCDA) that represents the collinear part,
and the decay constant that involves the interactions of scales $\lesssim m_Q$. By solving the renormalization-group (RG) equation for collinear part or the hard part, one can resum the logarithms $\ln r$.

In general, the collinear part describing a light meson such as a pion, $\rho$, or $\eta$ is nonperturbative and one usually introduces an LCDA with a few model parameters. However, in the case of heavy quarkonium, the collinear parts can further be factorized into perturbative short-distance coefficients at the scale $m_Q$ and nonperturbative long-distance matrix elements at the scale $m_Q v^2$ in the framework of NRQCD \cite{Jia:2008ep,Jia:2010fw,Wang:2013ywc,Bell:2008er}. Therefore, it is worth to revisit and to update predictions by including the resummation of the large logarithms at energies of future lepton colliders. We express our formula in such a way that our expression reproduces the fixed-order results at low energies $\sim m_Q$, and at higher energies it resums large logarithms so that the same expression can be used for both the Belle and future high-energy experiments. 

We consider the charge conjugate even ($C=+1$) processes with $S$-wave pseudoscalar quarkonium such as $\eta_{c}+ \gamma$ and $\eta_{b}+ \gamma$. In a fixed-order perturbation theory this process was first computed in \cite{Chung:2008km} at leading order (LO) and its next-to-leading order (NLO) correction was computed analytically in \cite{Sang:2009jc,Braguta:2010mf} and numerically in \cite{Li:2009ki}.  Up to date the $\alpha_s^2$ correction is available \cite{Feng:2015uha,Chen:2017pyi}. The relativistic correction of the order $v^2$ was first considered in \cite{Fan:2012dy} and $\alpha_s v^2$ correction was also obtained in  \cite{Xu:2014zra}. The virtual $Z$ contribution was computed up to $\alpha_s$ correction in \cite{Chen:2013mjb,Chen:2013itc}.

The leading-logarithmic (LL) accuracy resumming $\alpha_s^n \log^n r $ terms was first achieved in \cite{Jia:2008ep}. The quarkonium LCDAs at NLO were obtained by matching QCD onto NRQCD in \cite{Bell:2008er,Wang:2013ywc}. In Higgs or $Z$ boson decay into $J/\psi+\gamma$ processes \cite{Bodwin:2014bpa,Bodwin:2016edd,Bodwin:2017wdu,Bodwin:2017pzj}, the next-to-leading-logarithmic (NLL) accuracy resumming $\alpha_s^{n+1} \log^{n} r$ was achieved and the Abel-Pad\'e method which enables to handle divergences appearing in computing the relativistic correction to the rates, was developed as well. Using the method, we make the prediction for $\eta_{c,b}+\gamma$ in lepton colliders at NLL+NLO plus the leading $v^2$-correction accuracy.

The rest of the paper is organized as follows. In \sec{formula} we explain the theoretical formula to achieve NLL+NLO accuracy and provide all the ingredients for that order. Section \ref{sec:results} presents numerical results for the cross section and for the $Z$-boson decay rate into this process and
\sec{comparison} compares our result at the Belle energy to the previous results and to Belle's recent limit \cite{Jia:2018xsy}. We finally summarize in \sec{summary}.

\section{Theoretical formula} \label{sec:formula}

The LC approach allows us to capture and to resum all logarithmic terms (singular),
while non-logarithmic terms (nonsingular) can be computed by NRQCD fixed-order perturbation theory.
We can express our full cross section as a sum of singular and nonsingular parts as in \cite{Kang:2013nha,Kang:2014qba}.
\be\label{eq:sigma}
\sigma(r; \mu,\mu_0,\mu_\ns) 
=\sigma^\sing (r;\mu,\mu_0) +\sigma^\ns(r,\mu_\ns)\,,
\ee
where sing and ns in the superscripts and subscripts denote singular and nonsingular, respectively.
In the singular part the scattering amplitude is factorized into a hard scattering kernel, an LCDA, and an NRQCD LDME. Each of them depends on a relevant energy scale such as $\mu\sim\sqrt{s}$, or $\mu_0\sim m_Q$.\footnote{There is the  non-perturbative scale of the order $m_Q v^2$, which is not explicitly denoted because LDMEs is not evolved in practice but rather determined at the scale $m_Q$ or $2m_Q$ as in conventional NRQCD approach.}
The renormalization group (RG) evolution between those scales enables us to resum large logarithms appearing in the fixed-order cross section and details of the evolution will be presented in coming subsections. 
On the other hand, if we turn off the resummation by setting all the scales being the same, it reduces to the singular part of the fixed-order cross section: $\sigma^\text{fixed-sing}(r;\mu)=\sigma^\text{sing}(r;\mu,\mu).$
The singular part of fixed-order cross section is given by 
\be \label{eq:sigfixedsing}
\sigma^\text{fixed-sing}(r;\mu) =\lim_{r\to 0} \sigma^\text{fixed}(r;\mu)=\sigma_0 \left[ 1+\frac{\alpha_s C_F}{4\pi} c^\text{sing} +\langle v^2\rangle c^\text{sing}_{v^2} \right] 
\,,\ee
where the coefficients are
\footnote{
If one expresses the $m_P$ in $\sigma_0$ in terms of $m_Q$ and $v^2$, 
then one finds that
\[
2m_P
\left[1+\langle v^2\rangle c_{v^2}^\textrm{sing}\right]
=
4m_Q\sqrt{1+\langle v^2\rangle}
\left[1+\langle v^2\rangle c_{v^2}^\textrm{sing}\right]
\approx
\left[
2\sqrt{m_Q}
\left(1-\frac{5}{12}\langle v^2\rangle\right)
\right]^{2},
\]
which agrees with 
$d_s^{(0)}$ and $d_s^{(v^2)}$ in Eq. (2.26) 
of Ref.~\cite{Xu:2014zra}.}
\begin{eqnarray} \label{eq:c}
c^\text{sing}&=&-\frac{2}{3}\left[ (9-6\log 2) \log r +9 (3+\log^2 2 -3\log 2) +\pi^2 \right]
\,,\nn\\
c^\text{sing}_{v^2}&=& -\frac{4}{3}
\,.
\end{eqnarray}
The born cross section $\sigma_0$ is given by
\be\\ \label{eq:sig0} 
\sigma_0 = \frac{16\pi^2 \alpha^2(\sqrt{s})\alpha(0) e_Q^2 \tilde{e}_Q^2 m_P}{3s^2} \frac{\langle \cO_{1} \rangle_{P}}{m_Q^2}
\,,\ee
where $\alpha$ is the fine structure constant, $e_Q$ is the fractional charge of a heavy quark $Q$,
$m_P$ is the quarkonium mass, $m_Q$ is the heavy quark mass, $\langle \cO_{1} \rangle_{P}=\langle 0|\chi^{\dagger}\psi|P \rangle
\langle P|\psi^{\dagger}\chi|0 \rangle$ is the non-relativistically normalized LO NRQCD LDME of the production of $S$-wave pseudoscalar quarkonium $P$, and the relativistic correction 
to the LO NRQCD LDME,
$\langle v^2 \rangle$, is defined by  
\begin{eqnarray}
\langle v^{2} \rangle 
\equiv 
\frac{1}{m_Q^{2}}
\frac{\langle P|\psi^{\dagger}\left(-\frac{i}{2} \overleftrightarrow{\bm{D}}\right)^{2}\!\!\chi|0\rangle}
	{\langle P|\psi^{\dagger}\chi|0\rangle},
\end{eqnarray}
where $\bm{D}=\bm{\nabla}-ig_s\bm{A}$ is the spartial part of the gauge-covariant derivative
and
$\psi^{\dagger}  \overleftrightarrow{\bm{D}}  \chi\equiv
\psi^{\dagger}( \bm{D}\chi)-(\bm{D}\psi)^{\dagger}\chi$.
$\tilde{e}_Q^2$ is given in \eq{eq:teQ2}
which includes the effect of both of the virtual photon and $Z$ boson propagators.
To make our paper self-contained we also copy the fixed-order cross section in~\cite{Sang:2009jc} into \appx{fixedsig}.

The nonsingular part is defined by subtracting the fixed-order singular part from the fixed-order cross section as
\be \label{eq:ns}
\sigma^\ns(r;\mu)= \sigma^\text{fixed}-\sigma^\text{fixed-sing}=
-r\, \sigma_0\left[ 1+\frac{\alpha_s C_F}{4\pi} c^\text{ns} +\langle v^2\rangle c^\text{ns}_{v^2} \right] 
\,,\ee
where the coefficients are $c^\text{ns}=(c^\text{fixed}-c^\text{sing})/(-r)
\approx -14.9-4.8\log r+\log^2 r+O(r)$ and $c^\text{ns}_{v^2}=(c_{v^2}^\text{fixed}-c_{v^2}^\text{sing})/(-r)=-1/3$.
Note that we pull out a prefactor $-r$ in \eq{eq:ns} to imply the relative suppression of nonsingular part in small $r$ region but this correction is still important at the Belle energy.
The nonsingular part from the $Z$-boson contribution is about $4m_Q^2/m_Z^2$ and remains small near the resonance and we omit them in this paper.

Now the full cross section reproduces the ordinary fixed-order result when we turn off the RG evolution: $\sigma^\text{fixed}(r;\mu)=\sigma(r; \mu,\mu,\mu) $.  
Therefore, at the energies where $\log r\sim O(1)$ hence $\mu\sim \mu_0$, the full cross section is consistent with the fixed-order results and at higher energies where $|\log r| \gg 1$ and $\mu \gg \mu_0$, the resummation implemented in full cross section becomes effective. Therefore the formula in \eq{eq:sigma} gives correct results at wide range of energies of current and future colliders.
For the precise predictions for various energies including current and future colliders,
both of the resummation of large logarithms and the fixed-order computation should be improved equivalently.

Now let us discuss the amplitude for the singular part. 
In lepton collisions an initial lepton pair annihilates into virtual gauge bosons such as $\gamma^*$ or, $Z^*$,
then a pair of quark and anti-quark produced from the bosons turns into a bounded quarkonium state by emitting a photon. The scattering amplitude for a pseudoscalar quarkonium $P$ plus a final photon can be written as
\be\label{eq:M}
i\cM(e^+e^-\to\gamma^*/Z^*\to P\gamma)
= 
L_I^\mu\,  \langle  P(p)+\gamma(\epsilon,k)\vert J^\mu_{V} \vert 0\rangle
\,,\ee
where the index $I=\gam^*,Z^*$ represents the virtual bosons. The current has the vector and axial vector components: $J^\mu_V=\bar{\psi}\gamma^\mu \psi$ and $J^\mu_A=\bar{\psi}\gamma^\mu\gamma_5 \psi$. But the axial current does not contribute due to the opposite charge conjugation and we only have $J_V$ in \eq{eq:M}. 
 The part $L_I^\mu$ contains a matrix element of initial lepton pair, a virtual boson propagator, and electroweak charges of the quarks. Its expression is
\begin{eqnarray}
L_I^\mu
=
\begin{cases}
\displaystyle
\frac{ie_Qe^2}{s}\bar{v}(\bar{P}) \gamma^\mu u(P),
& \textrm{for $I=\gamma^*$},
\\[2ex]
\displaystyle
-\frac{ig_V^Q e^2/\sin^2(2\theta_W)}
{s-m_Z^2}\bar{v}(\bar{P}) \gamma^\mu(g_V^e-g_A^e\gamma_5) u(P),
& \textrm{for $I=Z^*$},
\end{cases}
\end{eqnarray}
where $e$ is the electromagnetic coupling, 
$e_Q$ is the fractional electric charge of the heavy quark $Q$,
 $\sqrt{s}$ is the CM collision energy,
$\theta_W$ is the Weinberg angle,
$g_V^e=-\frac{1}{2}+2\sin^2\theta_W$ and $g_A^e=-\frac{1}{2}$
are the vector and axial charges of electron,
and $g_V^Q=T_Q-2e_Q\sin^2\theta_W$ with $T_Q=\pm 1/2$
is the vector charge of quark with a flavor $Q=c,b$, respectively.

 The quark matrix element is extensively studied in the context of the meson form factor and the quark part in \eq{eq:M} can be expressed in the form factor style as
\bea\label{eq:Jmu}
\langle P(p)+\gamma(\epsilon,k)\vert J^\mu_{V} \vert 0\rangle
&=&-i\epsilon^*_\nu(k) ee_Q\int d^4 x e^{i k\cdot x}\,\left\langle P(p)\left\vert T\left[ J^{\nu \dagger}_{V}(x)J^\mu_{V}(0) \right]\right\vert 0\right\rangle
 \nn\\
 &=& i\epsilon_ \nu^{*}(k)\, ee_Q \frac{\epsilon_\perp^{\mu\nu}}{2} \, G_P(\mu) 
+ O(r)
\,,\eea
where $T$ is the time-ordered product and $\epsilon_\perp^{\mu\nu}=\epsilon^{\mu\nu\rho\sigma} p_\rho k_\sigma/(p\cdot k)$ is the asymmetric tensor.

\subsection{Light-cone distribution amplitude}

The factor $G_P(\mu)$ is the leading-twist result in LC factorization:
\begin{equation}
\label{eq:GPmu-def}
G_P(\mu)
\equiv
f_P(\mu,m_Q)
\int_0^1 dx \,
T_H^P(x;\mu,\sqrt{s})
\phi_P(x;\mu,m_Q),
\end{equation}
where $T_H^P$ is the hard-scattering kernel, $\phi_P$ is the LCDA of a pseudoscalar quarkonium $P$,
and $f_P$ is the decay constant. The scale $\mu$ is 
an arbitrary energy scale that separates the natural scales $m_Q$ and $\sqrt{s}$
which are the last arguments of each functions.
Each function depends on the logarithm of their ratio: the natural scale to the scale $\mu$. For simplicity, we omit the last arguments to the functions from now on.

The hard-scattering kernel $T_H^P(x;\mu)$ describes a production of quark and anti-quark pair at $^1S_0$ state. 
The one-loop expression is given in Ref.~\cite{Braaten:1982yp,Wang:2013ywc} by
\begin{equation}\label{eq:TH}
T_H^P(x,\mu)
=
T_H^{(0)}(x)
+
\frac{\alpha_s(\mu)}{4\pi}T_H^{(1)}(x,\mu)
+O(\alpha_s^2),
\end{equation}
where
\begin{subequations}
\begin{eqnarray}
T_H^{(0)}(x)&=&\frac{1}{\bar{x}}+(x\leftrightarrow \bar{x}),
\\
\label{eq:TH01}
T_H^{(1)}(x,\mu)&=&
\frac{C_F}{\bar{x}}
\left[
(3+2\log\bar{x})
\left(\log\frac{s}{\mu^2}-i\pi
\right)
+\log^2 \bar{x}+(8\Delta-1)\frac{\bar{x}\log{\bar{x}}}{x}-9
\right]+(x\leftrightarrow \bar{x}),
\phantom{moretabs}
\end{eqnarray}
\end{subequations}
here $\bar{x}\equiv 1-x$.
Note that $\Delta=0$ for the naive dimensional regularization (NDR) scheme \cite{Chanowitz:1979zu}
and $\Delta=1$ for the t'Hooft-Veltman (HV) scheme \cite{tHooft:1972tcz,Breitenlohner:1977hr} for $\gamma_5$ regularization.\footnote{ 
In the NDR scheme, $\{\gamma_5,\gamma^\mu\}=0$, for the index $\mu$ in $d$ dimension, while in the HV scheme, $\gamma_5$ defined in 4 dimension anticommutes  $\{\gamma^\mu,\gamma_5\}=0$ for $\mu=0,1,2,3$ but commutes $[\gamma^\mu,\gamma_5]=0$ for $\mu=4,\cdots, d$.
Note that the $\delta$ in Ref.~\cite{Braaten:1982yp}
and the $\Delta$ in Ref.~\cite{Wang:2013ywc} are related as $\delta=1-\Delta$.}
The scheme dependence in the hard kernel is cancelled by the same term with the opposite sign in the LCDA. 

The pseudoscalar LCDA is defined by a non-local matrix element of $\gamma^{+} \gamma_5$ as
\be
\label{def:LCDA-fP}
\langle P(p) | \bar{Q}(z) 
\gamma^{+} \gamma_5[z,0] Q(0) | 0\rangle
=
p^+\, f_P 
\int_0^1 dx\,e^{ip\cdot z x}\phi_P(x,\mu)\,, 
\ee
where the plus components are $\gamma^+=\gamma^0+\gamma^3$ and $p^+=p^0+p^3$.
$x\in[0,1]$ is the collinear momentum fraction of a quark in the quarkonium and $[z,0]$ is the gauge link that is defined by 
\be\label{eq:z0}
[z,0]=
\cP\text{exp}\left[i g_s \int_0^z dy\,A^+_a T^a \right]
\,,\ee
where $\cP$ stands for path ordering, 
$g_s=\sqrt{4\pi \alpha_s}$ is the strong coupling constant, 
$A^{a}$ is the gluon field with color index $a$, 
and $T^{a}$ is fundamental representation of SU$(N_c)$.

The LCDA $\phi_P$ describes the collinear-gluon exchange between quark
and anti-quark pair and it is normalized to the unity upon the integration over $x$. This normalization defines the decay constant $f_P$ to be \eq{def:LCDA-fP} at $z=0$ and it describes a $c\bar c$ pair transition into a physical quarkonium.\footnote{The definition of $f_P$ is different from that of Ref.~\cite{Wang:2013ywc} by a multiplicative factor $i$.}
In light mesons, the LCDA and the decay constant are nonperturbative and  the former is modeled with a few parameters and the latter is determined by comparison to measurement.
On the other hand in heavy quarkonium the LCDA and short-distance part of the decay constant are perturbatively calculable by matching QCD onto NRQCD amplitude. 
Their one-loop correction was obtained in \cite{Wang:2013ywc} and the relativistic correction was obtained in \cite{Bodwin:2014bpa,Wang:2017bgv}. We treat $v^2$ and $\as$ corrections are of the same size and expand up to the same power.
Then, the LCDA expanded up to the leading corrections in $\as$ and $v^2$ is
\be\label{eq:phiP}
\phi_P(x,\mu)
=
\phi^{(0)}(x)
+\frac{\alpha_s(\mu)}{4\pi} \phi^{(1)}
+\langle v^2 \rangle \phi^{(v^2)}
+O(\alpha_s^2, \alpha_sv^2,v^4),
\ee
where
\bea\label{eq:phi01}
\phi^{(0)}
&=&
\delta(x-\tfrac{1}{2}),
\nonumber \\
\phi^{(1)}
&=&
C_F\theta(1-2x)
\left\{
\left[
4x\frac{\tfrac12+\bar x}{\tfrac12-x}
\left(
\log\frac{\mu_0^2}{4m_Q^2}-2\log(\tfrac12-x)-1
\right)
\right]_+
+
\left[
\frac{4x\bar{x}}{(\tfrac12-x)^2}
\right]_{++}
+\Delta\left[16x\right]_+
\right\}
\nonumber\\&&
+(x\leftrightarrow \bar{x}),
\nonumber \\
\phi^{(v^2)}
&=&
\frac{\delta^{(2)}(x-\tfrac{1}{2})}{24},
\eea
and the $+$ and $++$ functions are defined in \appx{plus}.
The leading $\alpha_s$ and $v^2$ corrections to the decay constant $f_P$ are given by
\begin{equation}
\label{eq:fP}
f_P(\mu)
=
\frac{\sqrt{2N_c}\sqrt{2m_P}
\Psi_P(0)}{2m_Q}\left[1-\langle v^2\rangle
+\frac{\alpha_s(\mu)C_F}{4\pi}(-6+4\Delta)
+O(\alpha_s^2, v^4, \alpha_sv^2)\right]\,,
\end{equation}
where $\Psi_P(0)$ is the wavefunction at the origin  and is defined by $\Psi_P(0) = \langle P(p) | \psi^\dagger \chi | 0 \rangle/\sqrt{2N_c}$ and $|\Psi_P(0)|^2=\langle \cO_1\rangle_{P }/(2N_c) $.\footnote{We take the nonrelativistic normalization for the LDME while Ref.~\cite{Wang:2013ywc} takes the relativistic normalization (see Eq.~(4.8)): $\langle O(^1S_0)\rangle=\sqrt{2m_P}\langle P(p) | \psi^\dagger \chi |0\rangle$. }
The relativistic correction agrees with the result in Ref.~\cite{Wang:2017bgv} with $x_0=\bar{x}_0=1/2$.
The renormalon ambiguity coming from the pole mass 
can be avoided if we replace $m_Q$ by $\overline{\textrm{MS}}$
mass.\footnote{See, for example, Ref.~\cite{Bodwin:1998mn}.}
Thus we replace the pole mass $m_Q$ in \eq{eq:fP} with
the one-loop corrected $\overline{\textrm{MS}}$ mass 
in Ref.~\cite{Tarrach:1980up}:
$
m_Q=\overline{m}_Q
\left[
1+\as(\overline{m}_Q)C_F/\pi
\right]
$
and truncate higher-order contributions than our working precisions.
The singular part of the cross section can be written 
in terms of $G_P(\mu)$ in Eq.~(\ref{eq:GPmu-def}) as
\be\label{eq:sigma-GP}
\sigma^\text{sing}=\frac{2\pi^2e_Q^2 \tilde{e}_Q^2 
\alpha^2(\sqrt{s})\alpha(0)}{3s^2}
|G_P(\mu)|^2\,,
\ee
where for the virtual photon $\tilde{e}_Q^2$ is $e_Q^2$,  and for the virtual photon and $Z$ boson it is
\be\label{eq:teQ2}
\tilde{e}_Q^2 =
 e_Q^2 
-2\frac{e_Q g_V^Q g_V^e }
{\sin^2(2\theta_W)}
\frac{1-r_Z}
{(1-r_Z)^2
+r_Z\frac{\Gamma_Z^2}{s}}
+
\frac{  (g_V^Q)^2 [(g_V^e)^2+(g_A^e)^2]}
{\sin^4(2\theta_W)}
\frac{1}
{
(1-r_Z)^2
+r_Z\frac{\Gamma_Z^2}{s}
},
\end{equation}
where $r_Z=m_Z^2/s$. 
Note that \eq{eq:sigma-GP} is rather a fixed-order singular cross section in \eq{eq:sigfixedsing} because the functions \eqss{eq:TH}{eq:phiP}{eq:fP} are in fixed-order form. We obtained the resummed singular part after the RG evolution and resummation, which is discussed next subsection.

We also give the expression for the decay rate of $Z$ boson into a pseudoscalar quarkonium plus a photon in terms of $G_P(\mu)$ as
\be \label{eq:Gamma}
\Gamma^\text{fixed-sing}(r;\mu) =
\frac{\pi \alpha(m_Z)\alpha(0) e_Q^2 (g_V^Q)^2}
{6m_Z\sin^22\theta_W}
|G_P(\mu)|^2.
\ee

\subsection{RG equation and log resummation}
The large logarithms in the cross section can be resummed by
evolving each function in the factorization from its own natural scale, $\mu_0 \sim m_Q$ for the LCDA
or $\mu\sim \sqrt{s}$ for the hard-scattering kernel $T_H^P$ to a common scale $\tilde\mu$, which can be chosen to be an arbitrary scale between $\mu_0$ and $\mu$ because the $\tilde\mu$ dependence should be exactly cancelled when evolutions of all the functions are combined together. One of the simple and conventional choices is to set $\tilde\mu=\mu$ then, the LCDA is just evolved from $m_Q$ to $\mu$, while the hard-scattering kernel is treated as fixed-order function. 

The LCDA evolution is governed by the RG equation called
the Efremov-Radyushkin-Brodsky-Lepage (ERBL) equation \cite{Efremov:1979qk,Lepage:1980fj}:
\begin{equation}
\label{eq:BL-equation}
\mu^2\frac{\partial}{\partial \mu^2}
\left[f_P(\mu)\phi_P(x,\mu)\right]
=
\int_0^1 dy\, V(x,y;\alpha_s(\mu))
\left[f_P(\mu)\phi_P(y,\mu)\right],
\end{equation}
where the ERBL kernel $V(x,y;\alpha_s(\mu))$ for a pseudoscalar meson was extensively studied
in the pion form factor. 
In the case of quarkonium the product $f_P \phi_P$ is factorized into two parts LDME and short-distance coefficients as in \eqs{eq:fP}{eq:phi01} and the RG equation \eq{eq:BL-equation} can be expressed into two set of RG equations: one for LDMEs and the other for the coefficients.  The formal equation is evolved from LDME's natural scale $m_Qv^2$ to $\mu$ and the latter is from $\mu_0$ to $\mu$. This way would better fit to the philosophy of scale separation in effective field theory framework. However, LDME scale is nonperturbative and evolution from the scale would not work. Conventionally the LDMEs are determined at the scale $\mu_0$ rather  $m_Qv^2$. Then, we simply run both parts from $\mu_0$ to $\mu$ by using \eq{eq:BL-equation}.

The LL and NLL accuracies are achieved by solving the ERBL equation with one- and two-loop kernels respectively. The kernel is known up to two loops in the NDR scheme and we use the NDR results to achieve NLL accuracy.
\begin{equation}
\label{eq:Vperp-up-to-two-loop}
V(x,y;\alpha_s(\mu))
= \sum_{n=0}^\infty \left(\frac{\alpha_s(\mu)}{4\pi}\right)^{n+1}
V^{(n)}(x,y)
\,,
\end{equation}
where the one-loop coefficient is given by 
\begin{equation}
V^{(0)}(x,y)
=
2C_F
\left[
\frac{1-x}{1-y}\left(1+\frac{1}{x-y}\right)\theta(x-y)
+\frac{x}{y}\left(1+\frac{1}{y-x}\right)\theta(y-x)
\right]_+
\,,
\end{equation}
and the two-loop expression can be found in Refs.~\cite{Dittes:1983dy,Sarmadi:1982yg,Mikhailov:1984ii,Katz:1984gf}.
The eigenfunction of the one-loop kernel $V^{(0)}(x,y)$ is $G_n$
whose eigenvalue is $-\gamma_n^{(0)}/2$:
\be
\label{eq:V0-Gn-relation}
\int_0^1 dy\, V^{(0)}(x,y)G_n(y) =-\frac{\gamma_n^{(0)}  }{2}\,G_n(x),
\ee
where $G_n$ is the product of the Gegenbauer polynomial $C_n^{(3/2)}$ and its weight \cite{Jones:2007zd}:
\begin{equation}
G_n(x)=x(1-x)C_n^{(3/2)}(2x-1).
\end{equation}
$\gamma_n^{(0)}$ is the LO anomalous dimension and here we follow the convention of Ref.~\cite{Agaev:2010aq}
\be
\label{eq:anomalous-dim}
\gamma_n^{(0)}
=
8C_F\left[H_{n+1}-\frac{1}{2(n+1)(n+2)}-\frac{3}{4}\right].
\ee
The NLO anomalous dimension $\gamma_n^{(1)}$ is defined in the same way from the 2-loop kernel $V^{(1)}(x,y)$.
The solution of the ERBL equation is expressed as a series sum,
\be\label{eq:phin-mu}
\phi_n(\mu)
=\sum_{k=0}^n
U_{nk}(\mu,\mu_0)\phi_k(\mu_0)
\,,\ee
where the $k$-th Gegenbauer coefficient of LCDA at the scale $\mu_0$ is
\bea
\label{eq:phin-mu0}
\phi_n(\mu_0)
=N_n\int_0^1 dx\,
\phi_P(x,\mu_0)
C_n^{(3/2)}(2x-1)\,,
\eea
with $N_n= 4(2n+3)/[(n+1)(n+2)]$. The coefficient at LO is simple
\be\label{eq:phi0n}
\phi^{(0)}_n = N_n C^{(3/2)}_n(0)\,,
\ee
where $C^{(3/2)}_n(0) =(-1)^{n/2} \tfrac{(n+1)!!}{n!!}$ for even $n$ and zero for odd $n$.
Explicitly, the LO ERBL equation for the $n$-th moment of
the LCDA is given by
\begin{equation}
\label{eq:ERBL-LO}
\mu^2\frac{d}{d \mu^2}
\left[f_P(\mu)\phi_n(\mu)\right]
=
\frac{\alpha_s(\mu)}{4\pi}
\frac{(-\gamma_n^{(0)})}{2}
\left[f_P(\mu)\phi_n(\mu)\right].
\end{equation}

Following the convention of Ref.~\cite{Agaev:2010aq}, the solutions
of the scale evolution factor $U_{nk}(\mu,\mu_0)$ up to NLL accuracy are given by 
\begin{equation}
\label{eq:Unk}
U_{nk}(\mu,\mu_0)
=
\left[
\frac{\alpha_s(\mu)}{\alpha_s(\mu_0)}
\right]^{\frac{\gamma_n^{(0)}}{2\beta_0}}
\left[
\delta_{nk}
\left(
1+
\frac{\alpha_s(\mu)-\alpha_s(\mu_0)}{4\pi}
\frac{\gamma_n^{(1)}\beta_0
-\gamma_n^{(0)}\beta_1}{2\beta_0^2}
\right)
+
(1-\delta_{nk})
d_{nk}(\mu,\mu_0)
\frac{\alpha_s(\mu)}{4\pi}
\right],
\end{equation}
where the value of $U_{nk}$ at LL accuracy is nonzero only for $n=k$: $\left[\alpha_s(\mu)/\alpha_s(\mu_0)\right]^{\gamma_n^{(0)}/(2\beta_0)} \delta_{nk}$.
 At NLL it is nonzero when $(n-k)$ is zero or, even and positive integer.
 $\beta_n$ is the beta function coefficients for $(n+1)$-th order in $\as$
 and the explicit expressions of the two-loop anomalous dimensions $\gamma_n^{(1)}$ and $d_{nk}$ are 
copied in \appx{anomalous-dim}.

Then, the RG evolved function $G_P(\mu)$ is given by
\begin{eqnarray}
\label{eq:GP-NLL-final}
G_P(\mu)
&=&
f_P\,
\bigg\{
\cM^{(0,0)}(\mu)
+
\frac{\alpha_s(\mu)}{4\pi}
\cM^{(1,0)}(\mu)
+
\langle v^2\rangle_P
\cM^{(0,v^2)}(\mu)
+
\frac{\alpha_s(\mu_0)}{4\pi}
\cM^{(0,1)}(\mu)
\bigg\},
\phantom{xx}
\end{eqnarray}
where $\cM^{(i,j)}$ are defined in terms of the RG evolved LCDA in \eq{eq:phin-mu} 
\be
\label{eq:Mij}
\cM^{(i,j)}(\mu)= \sum_{n=0}^\infty T_n^{(i)}(\mu) \phi_n^{(j)}(\mu)
\,,
\ee
and 
\be
T_n^{(i)}(\mu) =\int_0^1 dx\, T_H^{(i)}(x,\mu) G_n(x)
\end{equation} is the coefficient in the expansion with Gegenbauer polynomials and it is non-vanishing only for even $n$ because $T_H^P$ is symmetric with respect to $x=1/2$ while $G_n(x)$ is asymmetric for odd $n$.
The LO coefficient for even $n$ is simple 
\be\label{eq:T0n}
T_n^{(0)}=1 
\,.\ee

We would like to note that the decay constant $f_P$ in \eq{eq:GP-NLL-final} is not evolved at NLL because its one- and two-loop anomalous dimensions in \eqs{eq:anomalous-dim}{eq:two-loop-anomalous-dim} are zeros:  $\gamma_0^{(0)}=\gamma_0^{(1)}=0$. One can see this explicitly by taking the 0-th Gegenbauer coefficient of $\phi_P$ in Eq.~(\ref{eq:BL-equation}),
or in Eq.~(\ref{eq:ERBL-LO})
because $\phi_0(\mu)=1$ for all $\mu$.
We also emphasize that the relativistic correction $\cM^{(0,v^2)}(\mu)$ correctly resums logarithms proportional to $\langle v^2\rangle \as^n \log^n r$
 by using the same RG evolution $U_{nk}$ and its expansion in $\as$ is given by
\be\label{eq:asv2}
 G_P\propto \langle v^2\rangle\left[ -\frac{5}{3}+ \frac{27-10\log 2}{3}\frac{\as(\mu_0^2) C_F}{4\pi}\log\frac{\mu_0^2}{\mu^2}\right]
\,. \ee
This agrees with the logarithmic term in $\as \langle v^2\rangle$ correction Eq.~(2.26) in \cite{Xu:2014zra}.

Eventually we insert \eq{eq:GP-NLL-final} into \eq{eq:sigma-GP} and obtain the singular part and full cross section in \eq{eq:sigma}.
In practice of computation there is an option of truncating higher-order terms irrelevant at our NLL accuracy and we make following truncations.  
In \eq{eq:GP-NLL-final}, the first term $\cM^{(0,0)}$ is computed using NLL expression of $U_{nk}$ while the other $\cM^{(i,j)}$ terms are computed using LL expression.  In the absolute square $|G_P(\mu)|^2$, we also drop higher-order terms proportional to $\as^2$ or $\as \langle v^2\rangle$, which are obtained in the product of 
$\cM^{(i,j)}$ in \eq{eq:GP-NLL-final} and $f_P$ in \eq{eq:fP}.

We also adopt the Abel-Pad\'e method developed and used in \cite{Bodwin:2016edd,Bodwin:2017pzj,Bodwin:2017wdu} to achieve faster numerical convergence at NLL accuracy and to deal with divergences associated with the relativistic corrections in LCDA.

\subsection{$\gamma_5$-scheme dependence}
As we can see from \eqss{eq:TH01}{eq:phi01}{eq:fP}, there are $\gamma_5$ scheme dependences in the hard kernel $T_H(x,\mu)$, LCDA $\phi_P(x,\mu)$ and decay constant $f_P(\mu)$, which are represented by the terms proportional $\Delta=0, 1$ for NDR and HV schemes, respectively. It is easy to check that the $\Delta$ dependences of the factor $G_P(\mu)$ vanish at NLO without resummation or, RG evolution. However, it is not obvious whether the $\Delta$ dependences vanish or not  at NLL accuracy due to additional scheme dependence that may enter in two-loop anomalous dimension $\gamma_n^{(1)}$. Note that $\gamma_n^{(0)}$ is $\Delta$-independent and so is the LL resummation. Ref.~\cite{Melic:2001wb} computed the $n_f$-dependent part of two-loop evolution kernel $V^{(1)}(x,y)$ in both NDR and HV schemes and we can obtain the scheme dependence for full two-loop evolution kernel by combining Eqs.~(5.24), (5.35), and (5.41a) and applying the relation in Eq.~(5.40) in Ref.~\cite{Melic:2001wb}, which gives
\begin{eqnarray} \label{eq:dV1}
\Delta V^{(1)}(x,y) = -8\Delta C_F\beta_0\left[\frac{x}{y}\theta(y-x)+\frac{1-x}{1-y}\theta(x-y)\right].
\end{eqnarray}
Again, polynomials $G_n(x)$ is the eigenfunction of $\Delta V^{(1)}$ with the eigenvalue anomalous dimension
\begin{eqnarray}\label{eq:dgam1-def}
\int_0^1 dy~\Delta V^{(1)}(x,y)G_n(y)= -\frac{\Delta\gamma_n^{(1)}}{2}G_n(x)\,,
\end{eqnarray}
where analytic expression of the anomalous dimension is given by 
\be\label{eq:dgam1}
\Delta\gamma_n^{(1)} =\Delta \frac{16 C_F \beta_0}{(n+1)(n+2)}\,.
\ee

At NLL, there are two types of $\Delta$ dependences in the amplitude $G_P(\mu)$. The one from NLL evolution factor $U_{nk}$ in \eq{eq:Unk} is proportional to the anomalous dimension in \eq{eq:dgam1}:
\be\label{eq:dan}
\Delta^{(a)}_{n} = G_{P,n}^\text{LL}(\mu)\frac{\alpha_s(\mu)-\alpha_s(\mu_0)}{4\pi}
\frac{\Delta \gamma_n^{(1)}}{2\beta_0}
\,.
\ee
Here $G_{P,n}^\text{LL}(\mu)=\left[
\tfrac{\alpha_s(\mu)}{\alpha_s(\mu_0)}
\right]^{\tfrac{\gamma_n^{(0)}}{2\beta_0}}\, f_P^{(0)}T_n^{(0)}\phi^{(0)}_n $ is the LL amplitude, where $f_P^{(0)}= \sqrt{N_c m_P}\Psi_P(0)/m_Q$ is LO decay constant and $T_n^{(0)}$, $\phi_n^{(0)}$ are given in \eqs{eq:phi0n}{eq:T0n}.

The other type of $\Delta$ dependence is those from one-loop corrections $f_P^{(1)}, T_n^{(1)}, \phi_n^{(1)}$.  The terms proportional to $\Delta$ are given by non-logarithmic constant parts
\begin{eqnarray}
\Delta f_P^{(1)}&=& 4C_Ff_P^{(0)}\, \Delta ,\\
\Delta T_n^{(1)}&=&  8C_F\Delta\int_0^1 dx \frac{1}{x\bar{x}}(x\ln{x}+\bar{x}\ln{\bar{x}})G_n(x),\\
\Delta\phi^{(1)}_n&=& 16C_F\Delta N_n \int_0^1 dx  \big[ \theta(1-2x) [ x ]_{+} + \theta(1-2\bar{x}) [\bar{x}]_{+} \big] C_n^{(3/2)}(2x-1).
\end{eqnarray}
Collecting three contributions above, we have
\begin{eqnarray}
\nn
\Delta^{(b)}_{n}
&=&G_{P,n}^\text{LL}(\mu)
\Big(\frac{\alpha_s(\mu_0)}{4\pi}\Delta f_P^{(1)}/f_P^{(0)}+\frac{\alpha_s(\mu_0)}{4\pi}\Delta\phi^{(1)}_n/\phi^{(0)}_n
+\frac{\alpha_s(\mu)}{4\pi}\Delta T_n^{(1)}/T_n^{(0)}\Big)
\\ \nn
&=& G_{P,n}^\text{LL}(\mu)
\Big(
\frac{\alpha_s(\mu)}{4\pi}\Big[\Delta f_P^{(1)}/f_P^{(0)}+\Delta\phi^{(1)}_n/\phi^{(0)}_n +\Delta T_n^{(1)}/T_n^{(0)} 
\Big]
\\ 
&&\qquad\qquad 
+\frac{\alpha_s(\mu_0)-\alpha_s(\mu)}{4\pi}\Big[\Delta f_P^{(1)}/f_P^{(0)} +\Delta\phi^{(1)}_n/\phi^{(0)}_n \Big]
\Big)
\,. \label{eq:mb}
\end{eqnarray}
In the second equality we rearranged terms into two parts, the one proportional to NLO correction evaluated at the $\as$ scale $\mu$  and the other proportional to the difference $\as(\mu_0)-\as(\mu)$ which is a part of NLL resummation. The scheme independence at fixed-order NLO implies the cancellation of first part
\be\label{eq:DeltaNLO}
\Delta f_P^{(1)}/f_P^{(0)}+\Delta\phi^{(1)}_n/\phi^{(0)}_n +\Delta T_n^{(1)}/T_n^{(0)} =0
\,.\ee
This is also confirmed by explicit computing $\Delta T_n^{(1)}$ and $\Delta\phi^{(1)}_n$ which are zero for odd $n$ and 
\begin{eqnarray}
\label{eq:dt1}
\Delta T_n^{(1)}&=&  -\frac{8C_F\Delta}{(n+1)(n+2)} ,\\
\label{eq:dphi1}
\Delta\phi^{(1)}_n&=&-4C_F \Delta N_n C^{(3/2)}_{n}(0) \left(1-\frac{2}{(n+1)(n+2)}\right)
\,,
\end{eqnarray}
for even $n$.
We would like to note an interesting relation between  $\Delta T_n^{(1)}$ and $\Delta\gamma_n^{(1)}$:
\be\label{eq:T1dg1}
\Delta T_n^{(1)}=-\Delta\gamma_n^{(1)}/(2\beta_0)
\ee 
or, $\Delta\phi^{(1)}_n/\phi^{(0)}_n+\Delta f_P^{(1)}/f_P^{(0)}=\Delta\gamma_n^{(1)}/(2\beta_0)$ equivalently. This implies that the constant term in the one-loop functions completely determines $\Delta$ dependence of two-loop evolution kernel and this ensures the cancellation of $\Delta^{(a)}_{n}$ and $\Delta^{(b)}_{n}$:
\be\label{eq:Deltasum}
\Delta^{(a)}_{n}+\Delta^{(b)}_{n}
=  G_{P,n}^\text{LL}(\mu)\frac{\alpha_s(\mu)-\alpha_s(\mu_0)}{4\pi}
\left[\frac{\Delta \gamma_n^{(1)}}{2\beta_0}+ \Delta T_n^{(1)}/T_n^{(0)}\right]=0
\,, 
\ee
where in $\Delta^{(b)}_{n}$ we eliminated $\Delta f_n^{(1)}$ and $\Delta \phi_n^{(1)}$ in favor of $\Delta T_n^{(1)}$ by using \eq{eq:DeltaNLO}.
Therefore, $\gamma_5$ scheme independence is valid at NLL accuracy.

At higher-order, we expect a similar pattern of cancellation between $\Delta$ dependent terms: cancellation between fixed-order terms at the same $\as$ scale as in \eq{eq:DeltaNLO} and cancellations between $n$-loop anomalous dimension from evolution factor and the constant terms of $(n-1)$-loop function as in \eq{eq:Deltasum}.

\subsection{Logarithmic structure}
Here we discuss logarithmic structure and accuracy of the resummed amplitude. Even though this section explains quite well-known properties of resummation and does not contain anything new, it may be useful for those who are not familiar with resummation.

Let us first look at the fixed-order expansion of amplitude in \eq{eq:GPmu-def}. Its logarithmic structure can be schematically expressed as
\begin{align}
\label{eq:GPfo}
G_P^\text{fixed} = \    c_{00} \ 
 &+ \frac{\as}{4\pi}( c_{11}L + c_{10} )  \\
&+  \left(\frac{\as}{4\pi}\right)^2( c_{22} L^2 + c_{21}L + c_{20} ) 
+  \cdots
  \nn \,,
\end{align}
where $L\equiv \log (4m_Q^2/s)$. The largest logarithmic term at each $\as$ order is $\as^n L^{n}$ and then the next largest is $\as^{n} L^{n-1}$.
The functions $T_H$ and $\phi$ are first expanded with the Gegenbauer polynomials, then each coefficient is resummed as $G_P^{nk}=f_P \,T_n U_{nk} \phi_k$, and summation over all $n$ and $k$ gives the resummed $G_P$. The individual $G_P^{nk}$ takes the following form
\begin{align} \label{eq:GPnk}
G_P^{nk}
= C(\as) \ \exp\ \biggl[ \ \frac{\as}{4\pi}( &  C_{11}L +C_{10})   \\
  +  \left(\frac{\as}{4\pi}\right)^2(&C_{22} L^2 +   C_{21}L +C_{20})  + \cdots\biggr] 
  \nn\\ 
  &\text{LL \mbox{   }   }+\text{NLL\mbox{  }}+\cdots
\nn \,,
\end{align}
where $ C(\as)$ is the fixed-order expansion in $\as$ and it does not depend on the logarithms,
\be\label{eq:Cas}
C(\as) = C_0 +\frac{\as}{4\pi} C_1+\cdots
\,. \ee
The fixed-order coefficient $C_i$ is given by fixed-order function $T_n$, $\phi_k$, $f_P$ and the coefficients $C_{ij}$ associated with anomalous dimensions are given by $U_{nk}$ in \eq{eq:Unk}. 
Therefore, in general the coefficients $C_i$ and $C_{ij}$ differ for the different values of $n,k$.
For example, $C_0$ is non-zero for the diagonal element where $n=k$ but zero otherwise.
We are implicit with those $n,k$ dependence to make our discussion focused on the logarithmic structure.
Similarly, we do not separately discuss about $v^2$ corrections in the coefficients $C_i$ and it follows the same conclusion.

In the fixed-order perturbation theory, the series in \eq{eq:GPfo} are summed row-by-row, i.e., order-by-order in $\as$.
On the other hand, in resummed perturbation theory, the series in the exponent of \eq{eq:GPnk} are summed column-by-column,
based on large-logarithmic power counting $L\sim 1/\as$.
In \eq{eq:GPnk}, the first column is of the order $\as^n L^n \sim 1$ called the LL,
the second column is $\as^{n} L^{n-1} \sim \as$ called the NLL, and so on.
It is clear that which fixed-order terms in \eq{eq:Cas} should be included: $C_0$ at LL and $C_{1}$ at NLL.

However, one may realize that the structure of evolution factor $U_{nk}$ in \eq{eq:Unk} is different from that of \eq{eq:GPnk}.
For example, the non-exponent term contains logarithms: $\as(\mu)-\as(\mu_0)= -\tfrac{\as(\mu_0)^2}{2\pi}\beta_0 \log(\mu/\mu_0) +\cdots$.
This is because,  in \eq{eq:GPnk} the second column is of $O(\as)$  hence those terms beyond LL can be expanded and moved down from the exponent:
\bea \label{eq:GPnk2}
G_P^{nk}&=& C(\as) 
\left(  1+   \frac{\as}{4\pi} C_{10} +\left(\frac{\as}{4\pi}\right)^2 C_{21}L+\cdots  \right)
\ \exp\ \biggl[ \ \frac{\as}{4\pi}   C_{11}+ \left(\frac{\as}{4\pi}\right)^2 C_{22} L^2   + \cdots\biggr] 
\nonumber
\\
&=&\tilde{C}(\as) \  \exp\ \biggl[ \ \frac{\as}{4\pi}   C_{11}+ \left(\frac{\as}{4\pi}\right)^2 C_{22} L^2   + \cdots\biggr] 
 \,,\eea
where $\tilde{C}(\as)$ includes two prefactors 
\be  \label{eq:tC}
\tilde{C}(\as) =C_0+\frac{\as}{4\pi} \left[ C_1+C_0 \sum_{n=0} C_{1n}\left(\frac{\as L}{4\pi}\right)^n \right]+ \cdots
\,.\ee
We have LL accuracy with first term in \eq{eq:tC} and NLL with $O(\as)$ terms in the large-logarithmic power counting $\as^{n+1} L^n \sim \as$.
This alternative way of arranging logarithms is equivalent to \eq{eq:GPnk} up to higher-order corrections than working accuracy and is the formula we use in this paper.

\section{Numerical results}\label{sec:results}

In this section, we list input parameters for numerical calculations then, present our results for  the final state $\eta_{c,b}+\gamma$ in $e^+e^-$ collisions at various collision energies and in $Z$-boson decay. Those results include the resummation at NLL accuracy, the fixed-order correction at NLO, and the relativistic corrections of the order $v^2$ as we discussed in previous sections.
The numerical results at LL and NLL+NLO are compared and their perturbative convergence is discussed.

\subsection{Input parameters and NRQCD matrix elements}

We use PDG values for $\overline{\text{MS}}$ mass $\msm_c=1.275^{+0.025}_{-0.035}~\text{GeV}$
 and $\msm_b=4.18^{+0.04}_{-0.03}~\text{GeV}$, which gives the one-loop pole mass
$m_c=1.483^{+0.029}_{-0.041}~\text{GeV}$ and $m_b=4.58^{+0.04}_{-0.03}~\text{GeV}$
and for Z-boson mass and width $m_Z=91.1876\pm 0.0021~\text{GeV}$ and $\Gamma_Z=2.4952\pm 0.0023~\text{GeV}$.
We run the coupling constants for the electroweak using the code Global Analysis of Particle Properties 
(GAPP) \cite{Erler:1998sy,Erler:1999ug}
and the coupling constants for the strong interaction 
using the 4-loop expression of the QCD beta function \cite{vanRitbergen:1997va}.
The CM energies of $B$-, $Z$- and Higgs factories are  $\sqrt{s}=10.58,\, 91.19,\, 240$\,GeV and
the values of coupling at respective energies are
$\alpha^{-1}(\sqrt{s})= 130.855,\, 127.916,\, 127.473$, $\sin\theta_W(\sqrt{s})= 0.233543,\, 0.231201,\, 0.236168$, and $\alpha_s(\sqrt{s})=0.1768,\, 0.1184,\, 0.1033$.

The NRQCD matrix elements such as the wave function at the origin and relative velocity were determined in \cite{Bodwin:2007fz} by using two constraints: electromagnetic decay rate $\Gamma [\eta_c\to \gamma\gamma ]$ of order $\as$ and the potential model. 
These values need updates due to changes in input parameters: charm-quark pole mass, scale of $\as$ from $m_{\eta_c}/2$ to $2 \overline{m}_c$, experimental value of the decay rate.
In the determinations of $\langle \cO_1\rangle_{\eta_c}$ and $\langle v^2 \rangle_{\eta_c}$, 
we used the same string tension $\sigma=0.1682\pm 0.0053~\text{GeV}^2$ \cite{Bodwin:2007fz} 
and updated values for the 1-loop pole mass  $m_c=1.483^{+0.029}_{-0.041}~\text{GeV}$, 
the mass difference between $J/\psi$ and $\psi(2S)$ $m_{2S}-m_{1S}=589.188\pm 0.028~\text{MeV}$, 
and the decay rate $\Gamma[\eta_c\to \gamma \gamma]=5.0\pm 0.4~\text{keV}$.\footnote{This is $30\%$ smaller than the value $7.2\pm0.7 \pm2.0$ keV used in \cite{Bodwin:2007fz}.}
Differently from \cite{Bodwin:2007fz} , we do not take average with $\langle \cO_1\rangle_{J/\psi}$
in the determination of $\langle \cO_1\rangle_{\eta_c}$.
In the decay rate formula, we have set the scale $\mu=2\msm_c$. 
The updated values are as follow:
\begin{eqnarray}\label{eq:MEetac}
\langle \cO_1\rangle_{\eta_c}&=&0.302^{+0.052}_{-0.049}~\textrm{GeV}^3,
\\
\langle v^2 \rangle_{\eta_c}
&=&
0.222^{+0.070}_{-0.070}.
\end{eqnarray}
The uncertainty  includes variations of $\sigma$, $m_c$, $m_{2S}-m_{1S}$, $\Gamma[\eta_c\to \gamma \gamma]$. 
And we assumed the size of the neglected higher-order corrections in $\alpha_s$ and $v^2$ to be $30\%$ times the central values of $\alpha_s$ and $v^2$, respectively. 
Major sources of uncertainty are the variation of $\sigma$ and the assumed higher-order corrections.

We like to pay a bit more attention to using those values in \eq{eq:MEetac}.
In conventional predictions, we may use the same value for the predictions at LO, NLO, and higher-order accuracy.
This way can correctly reproduce the input decay rate using its 1-loop expression, while with LO expression of the decay rate the result is systematically biased by the amount of 1-loop correction included in the matrix element.
Of course it is not a problem when the size of 1-loop correction is small as in bottomonium.
However in the case of $\eta_c$ we find the effect is as large as $40\%$ due to large coupling constant $\as(2\msm_c)\sim 0.27$.
 A similar bias exists  in the cross sections and leads to overshooting in its predictions at LO.
 Eventually it would spoil the perturbative convergence due to a large change from LO to NLO  and similarly from LL to NLL.

We can avoid this systematic bias once we use the matrix element determined at the same order with working order, at which we make predictions. By doing this the (experimental) input decay rate is always reproduced at each order in $\as$. This can be done by  replacing the NRQCD matrix element with the experimental value of decay rate multiplied by short distance coefficient:
\be\label{eq:OtoGam}
\langle \cO_1\rangle_{\eta_{Q}}= \Gamma_\text{exp} \frac{m_Q^2}{2\pi\alpha^2(0) e_Q^4}
\left[ 1 +\frac{4}{3} \langle v^2\rangle+ \frac{\as (\mu)  C_F}{\pi} \frac{20-\pi^2}{4}\right],
\ee
where the experimental value for $\eta_c$ is $\Gamma_\text{exp}=5.0\pm 0.4$~keV \cite{Tanabashi:2018oca}.
We insert \eq{eq:OtoGam} into \eq{eq:fP} and truncate higher-order terms than the working order.
One of advantages using \eq{eq:OtoGam} is that the error propagations associated with the pole mass and 
the NRQCD matrix element $\langle \cO_1\rangle_{\eta_c}$ become simpler. The pole mass $m_Q^2$ are cancelled by that of \eq{eq:sig0} in the cross section. 
The perturbative uncertainty obtained from scale variation of \eq{eq:OtoGam} largely contributes to uncertainty of $\langle \cO_1\rangle_{\eta_c}$ and 
this contribution is now naturally combined  in a correlated way with scale variations of the other part in the cross section.
Another advantage from an empirical observation is that the 1-loop correction in the decay constant reduces significantly due to the large cancellation between $O(\as)$ terms
of the decay constant \eq{eq:fP} and the decay rate in \eq{eq:OtoGam} as
\bea
\label{eq:fP-final}
|f_{\eta_{Q}}(\mu)|^2 &=&
\frac{2m_P \langle \cO_1\rangle_{\eta_{Q}} } {4m_Q^2}
\left[1-2 \langle v^2\rangle
+\frac{\alpha_s(\mu)C_F}{\pi}(-3) \right]\,
\nn\\
 &=&  \Gamma_\text{exp} [\eta_{Q}\to \gamma\gamma ]\frac{m_P}{4\pi\alpha^2(0) e_Q^4}\left[ 1  + \frac{\as(\mu) C_F}{\pi}  \frac{8-\pi^2}{4} 
-\frac{2}{3}\langle v^2\rangle \right]
\,.\eea
Note that  the coefficient of $\as C_F/\pi$ reduces from $-3$ to $(8-\pi^2)/4 \approx -0.5$ and the coefficient of 
$\langle v^2\rangle$ changes from $-2$ to $-2/3$.
In this way, we have better perturbative convergence between LL and NLL (LO and NLO). 

\subsection{Final results}

\begin{figure}[t]
\centering
\includegraphics[width=0.45\linewidth]{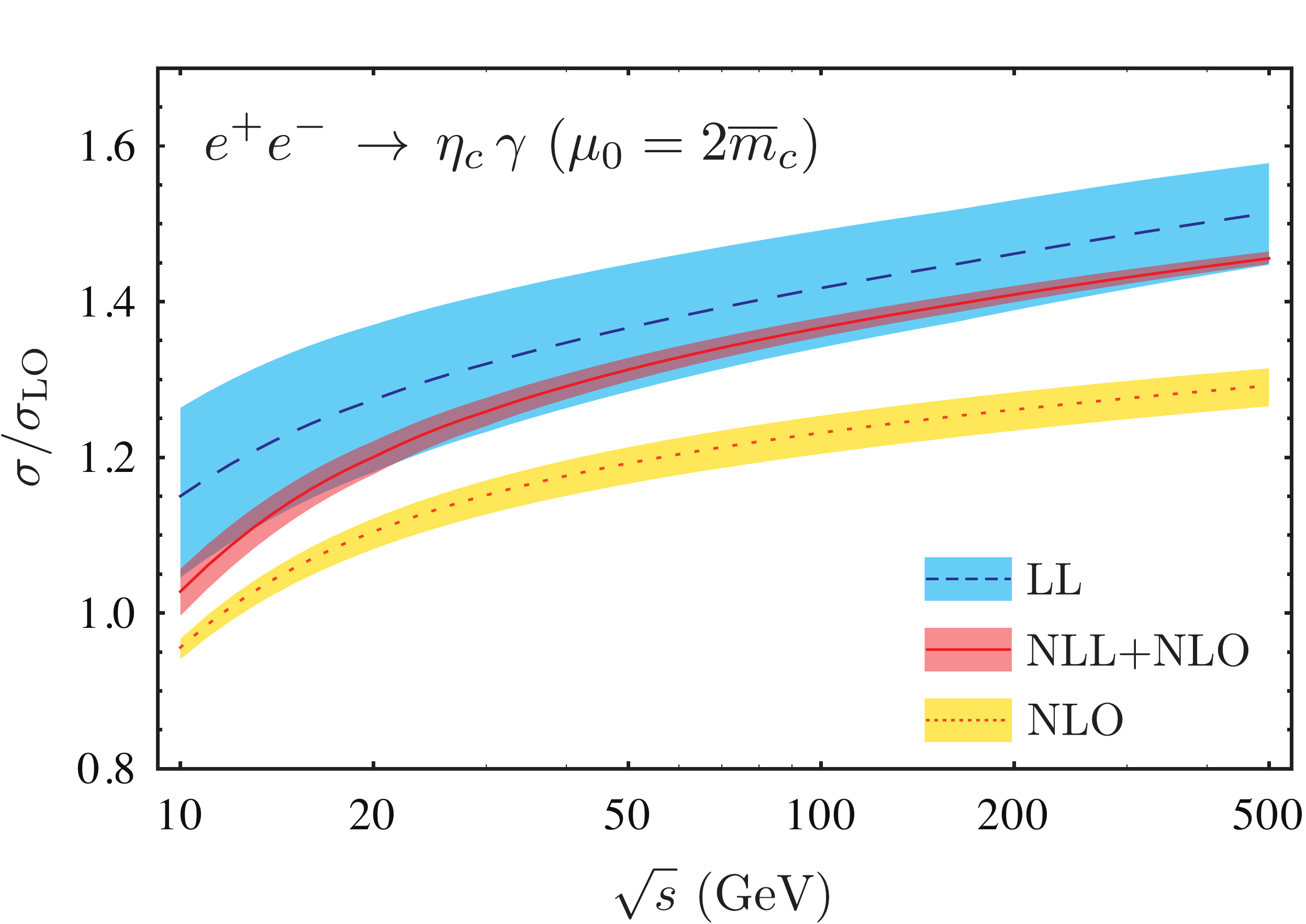}
\includegraphics[width=0.45\linewidth]{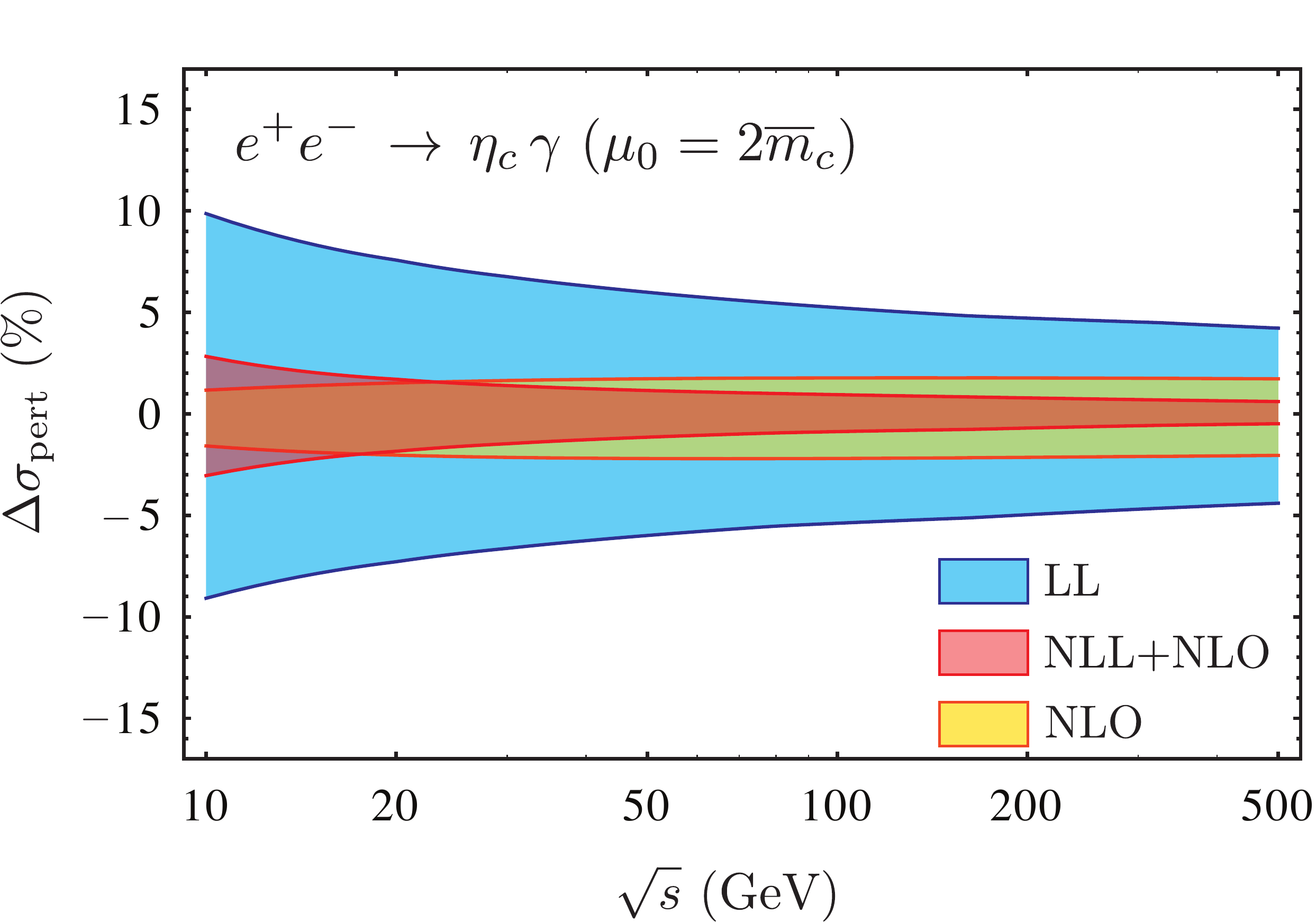}
\caption{
left panel: LL, NLL+NLO and NLO cross sections normalized by the LO cross section. The scale choices are
$\mu_0=2\msm_c$ and $\mu=\sqrt{s}$. Bands are perturbative uncertainties only.
right panel: perturbative uncertainties in percentage
\label{fig:charm-2mc}}
\end{figure}
\begin{figure}[tb]
\centering
\includegraphics[width=0.45\linewidth]{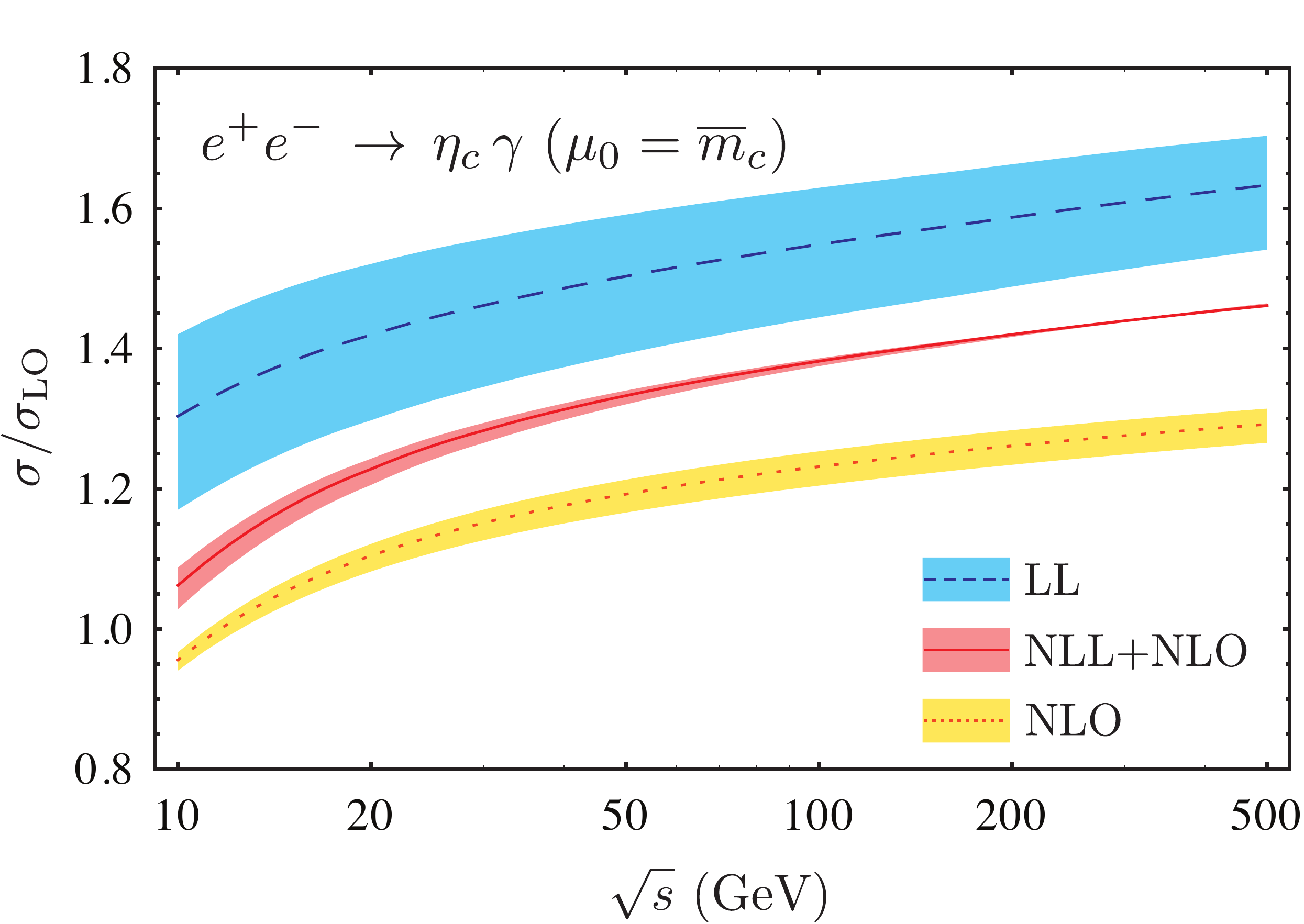}
\includegraphics[width=0.45\linewidth]{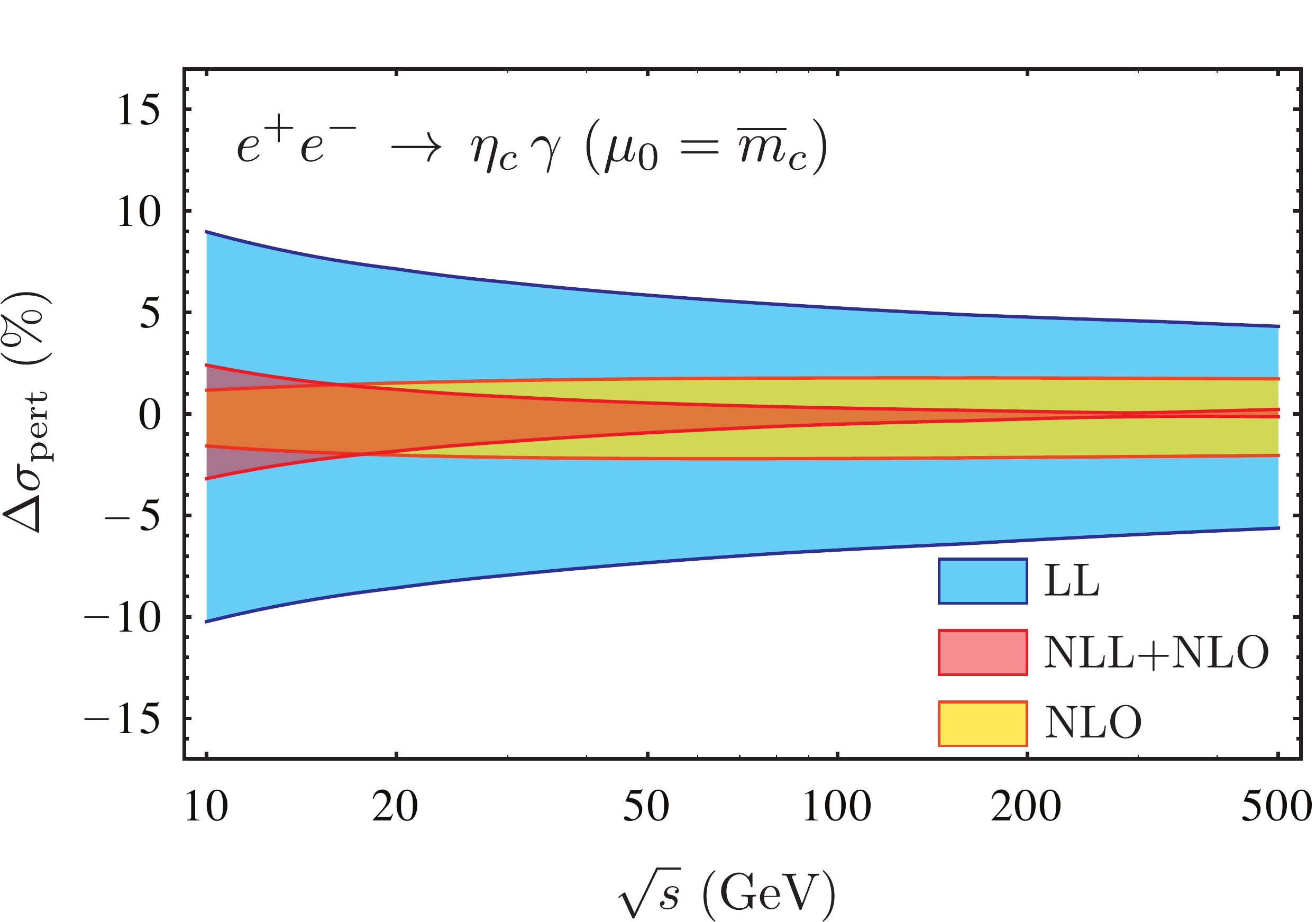}
\caption{
The same results with \fig{fig:charm-2mc} except for $\mu_0=\msm_c$
\label{fig:charm-mc}}
\end{figure}

Our numerical results for $\eta_c +\gamma$ cross sections and perturbative uncertainties at different accuracies are given in \fig{fig:charm-2mc}. Three accuracies LL, NLL combined with NLO non-singular part and leading $v^2$ correction (NLL+NLO), fixed-order NLO are compared. The bands on left and right panels are absolute and relative perturbative uncertainties. The cross section in figures is scaled by the LO cross section
\be\label{eq:LO}
\sigma_\text{LO}= \frac{8\pi \alpha^2(\sqrt{s})\tilde{e}_Q^2   m_P }{3\alpha(0)e_Q^2 s^2}\Gamma_\text{exp}
\,.
\ee
 The values of scales we choose for LL and NLL+NLO are $\mu_0=2\msm_c$ and $\mu=\mu_\text{ns}=\sqrt{s}$  and  for NLO we set all scales to be the same $\mu_0=\mu=\mu_\text{ns}=\sqrt{s}$.
 The perturbative uncertainties are estimated by varying $\mu$, $\mu_\text{ns}$ from its central value by a factor 2 up and down and by varying $\mu_0$ by a factor of $\sqrt{2}$. The uncertainties are summed in quadratures as in \cite{Kang:2013nha}: $\sqrt{\delta\sigma_{\mu_0}^2+\delta\sigma_{\mu}^2+\delta\sigma_{\mu_\text{ns}}^2}$, where $\delta\sigma_{\mu_i}$ is the change of cross section by a scale variation of $\mu_i$. Here we do not include other sources of uncertainty to show the perturbative convergence and they will be included later in the final results in \tab{tab:sigmas}.
The perturbative uncertainty (width of the band) decreases by a factor of $\as$ from LL to NLL+NLO and a reasonable overlapping between two bands in left panel implies a good perturbative convergence.
With increasing CM energy, the deviation of NLO from NLL+NLO becomes more significant due to the large logarithms not taken into account at NLO and this clearly shows that the small perturbative uncertainty of NLO is not reliable at this high energies.

In \fig{fig:charm-mc} we also show the results with a smaller value of $\mu_0$: $\mu_0=\msm_c$ instead of $2\,\msm_c$. 
The NLL perturbative uncertainty at $\mu_0=\msm_c$ tends to be asymmetric and smaller than that for $\mu_0=2\msm_c$ because the lower scale variation from $\msm_c$ by a factor of $\sqrt{2}$ moves $\mu_0$ close to the Landau pole and the scale dependence near this region is not monotonic.
In comparison to \fig{fig:charm-2mc}  at $\mu_0=2\msm_c$ we observe relatively better perturbative convergence between LL and NLL although the other is still reasonable. For these reasons we take $\mu_0=2\msm_c$ for our final results listed in \tab{tab:sigmas}.

\begin{figure}[t]
\centering
\includegraphics[width=0.47\linewidth]{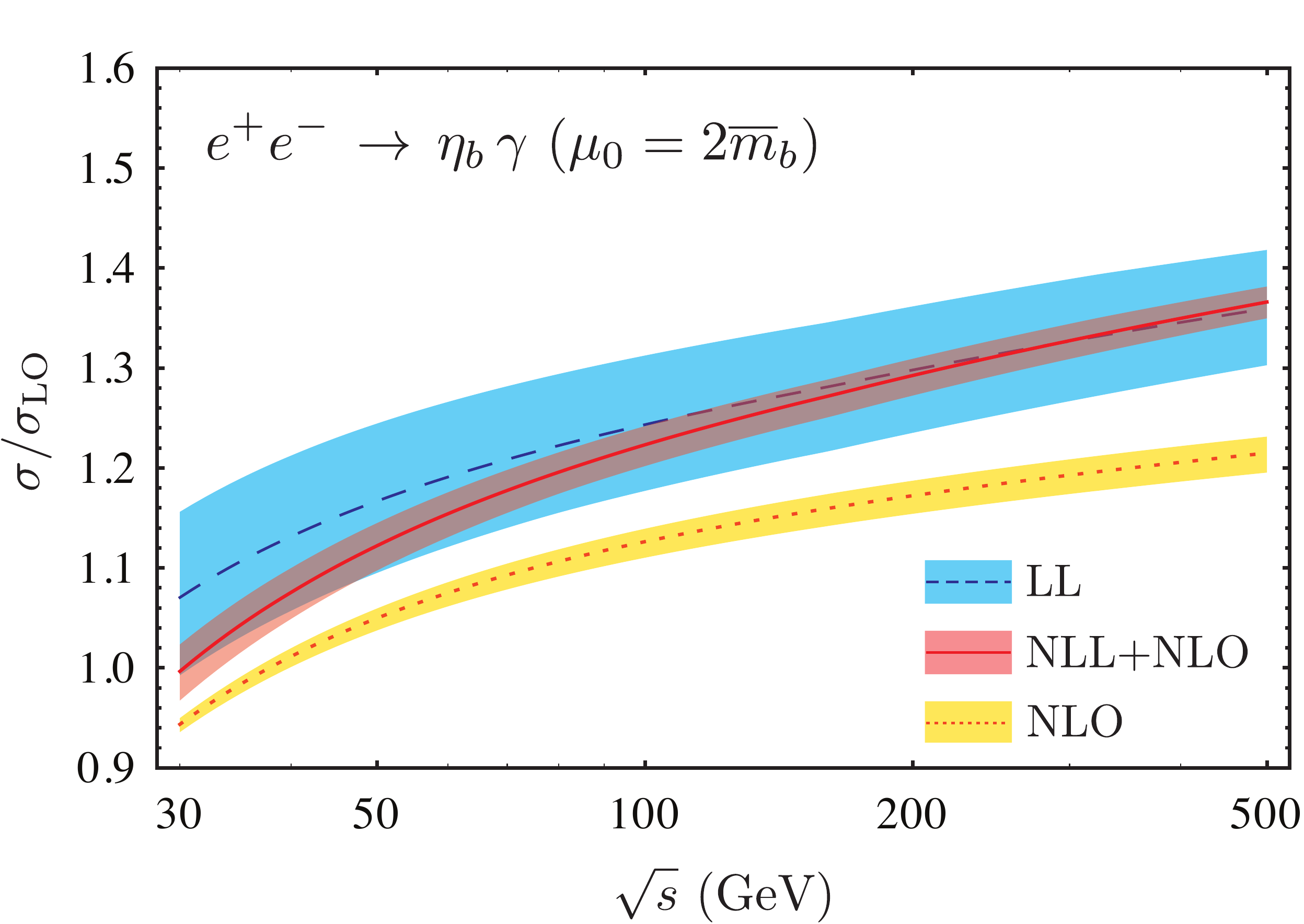}
\includegraphics[width=0.47\linewidth]{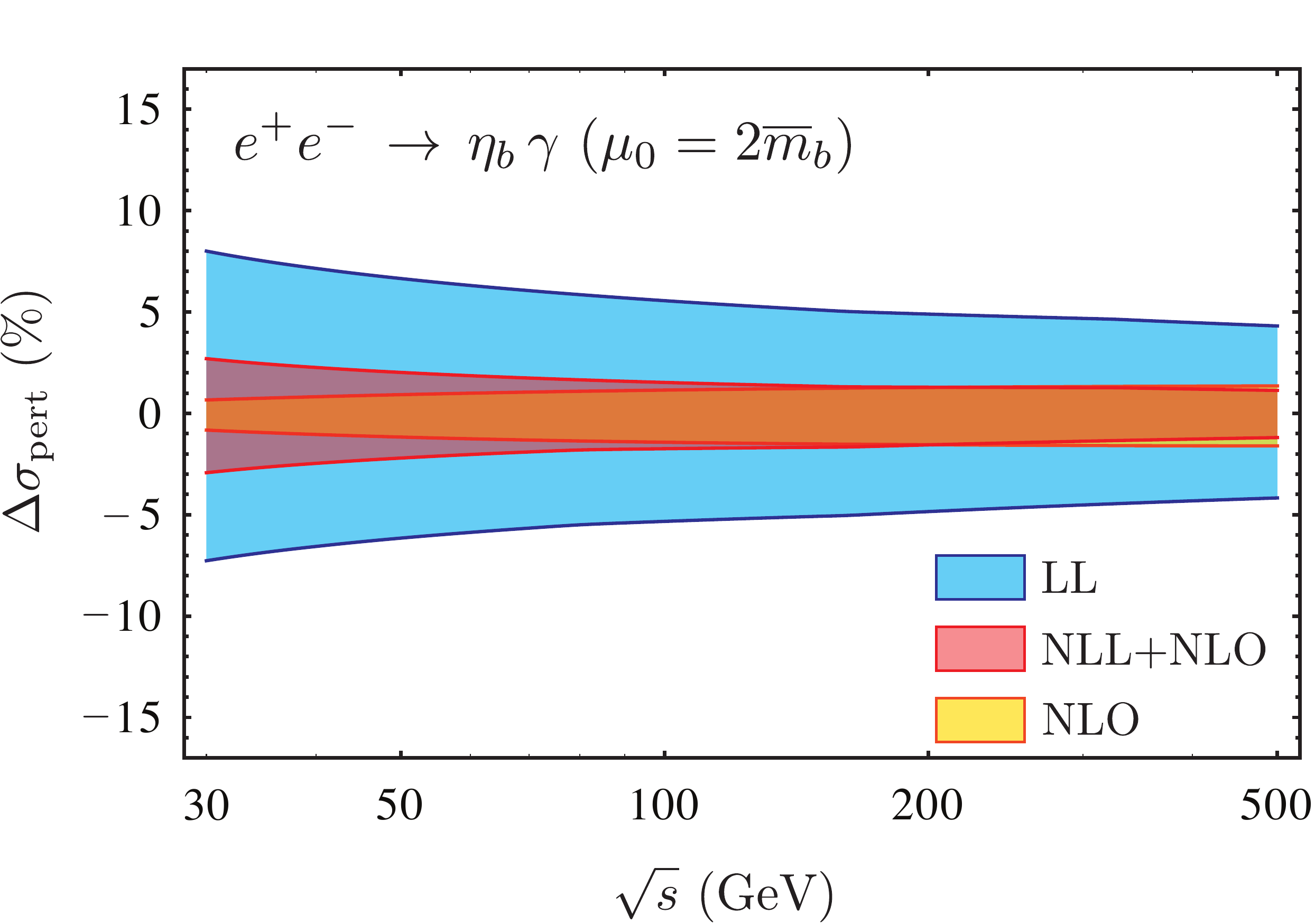}
\caption{
left panel: LL, NLL+NLO and NLO cross sections normalized by LO cross section with $\mu_0=2\msm_b$. 
right panel: corresponding perturbative uncertainties in percentage
\label{fig:bottom-2mb}}
\end{figure}

In \fig{fig:bottom-2mb} we also show the bottomonium production cross sections and their percent perturbative uncertainties, which are smaller compared those for charmonium. 
The decay rate for $\eta_b$ is not available we use following value
$\Gamma[\eta_b\to\gamma\gamma]=0.512^{+0.096}_{-0.094}~\text{keV}$
and for relative velocity $\langle v^2\rangle_{\eta_b}=-0.009^{+0.003}_{-0.003}$ taken from 
\cite{Chung:2010vz} \footnote{Note that there is no restriction to the positive definiteness of the matrix element $\langle v^2\rangle$.
The matrix element $\langle v^2\rangle$ intrinsically contains a linear ultraviolet (UV) divergence 
that must be regulated. For example, if we employ dimensional regularization that is consistent with
existing calulations of quarkonum decay and production rates at relative order $\alpha_s$ and $\alpha^2_s$,
then the scaless power divergent integrals are discarded. 
Such subtractions of divergent contributions can lead 
to both positive and negative values. See
Ref.~\cite{Bodwin:2006dn} for more details.}.
The central values of scales are $\mu_0=2\msm_b$ and $\mu=\mu_\text{ns}=\sqrt{s}$ and their variations are done in same way as for the charmonium.

\begin{table}[bth]
\begin{center}
\begin{tabular}{c|cc|cc}
\hline\hline
& \multicolumn{2}{c|}{Cross section}  &\multicolumn{2}{c}{Branching fraction}  
\\[-1ex]
$ \sqrt{s}$ & $ \eta_{c}$ & $\eta_{b}$ & $\eta_{c} $ & $\eta_{b}$
\\ \hline 
 10.58~GeV  &  $32.7\pm 2.8$ fb&  - & \multirow{3}{*}{$(7.42 \pm 0.61)\times10^{-9}$} &\multirow{3}{*}{$(2.80\pm 0.53)\times 10^{-8}$}\\  
 $m_Z$   		& $0.449\pm 0.037$ fb &  $1.66\pm{0.31}$ fb& \\ 
  240~GeV	&  $0.189 \pm 0.016$ ab & $0.0934\pm{0.0176}$ ab & \\ \hline \hline
\end{tabular}
\caption{Cross sections $\sigma(e^+e^-\to \eta_Q+\gam )$ and branching fractions Br($Z\to\eta_Q+\gam$) for charmonium and for bottomonium with uncertainties including all input parameters as well as perturbative uncertainties.
\label{tab:sigmas}}
\end{center}
\end{table}

\tab{tab:sigmas} lists our final results for the cross sections at $B$-, $Z$- and Higgs-factory energies: $\sqrt{s}=$10.58, 91.2, and 240~GeV and $Z$-boson decay branching fractions. The uncertainties in the table for charmonium ($\eta_c$) channel includes uncertainties of input decay rate $\Gamma^\text{exp}$ ($\pm 8\%$), relative velocity $\langle v^2\rangle$ ($\pm 30\%$) as well as perturbative uncertainties ($\pm 3\%$ or less) shown \fig{fig:charm-2mc} and they are added in quadratures.
For the bottomonium ($\eta_b$) the uncertainties are input decay rate ($\pm 20\%$), relative velocity ($\pm 30\%)$,  perturbation ($\pm 3\%$ or less).  The final uncertainties quoted in \tab{tab:sigmas} are dominated by uncertainty of input decay rate.\footnote{We do not include relatively small uncertainties from $\MSbar$ mass ($\pm 2\%$) and from higher-order electroweak corrections.}

There is an independent 
prediction for $\langle v^2\rangle_{\eta_b}=0.042$
in Ref.~\cite{Li:2012rn}. 
The authors of that reference 
have determined that value by making use of the Gremm-Kapustin relation.
If we use this numerical value to compute the branching fraction 
for $e^+e^-\to\eta_b+\gamma$ at $\sqrt{s}=m_Z$ given in 
Table~\ref{tab:sigmas}, we obtain $2.78\times10^{-8}$, 
which is well within
the prediction $(2.80\pm 0.53)\times 10^{-8}$ given 
in Table~\ref{tab:sigmas}.
However, we have not included the corresponding analysis 
into our final results listed in Table~\ref{tab:sigmas} because 
the determination of $\langle v^2\rangle$ by making use of
either the lattice or Gremm-Kapustin approaches suffers
too large uncertainties to determine 
even the signs of the matrix elements as is 
stated in Ref.~\cite{Bodwin:2006dn}.

\section{Comparison to various predictions and Belle's upper limit} \label{sec:comparison}

\begin{figure}[htb]
\centering
\includegraphics[width=0.6\linewidth]{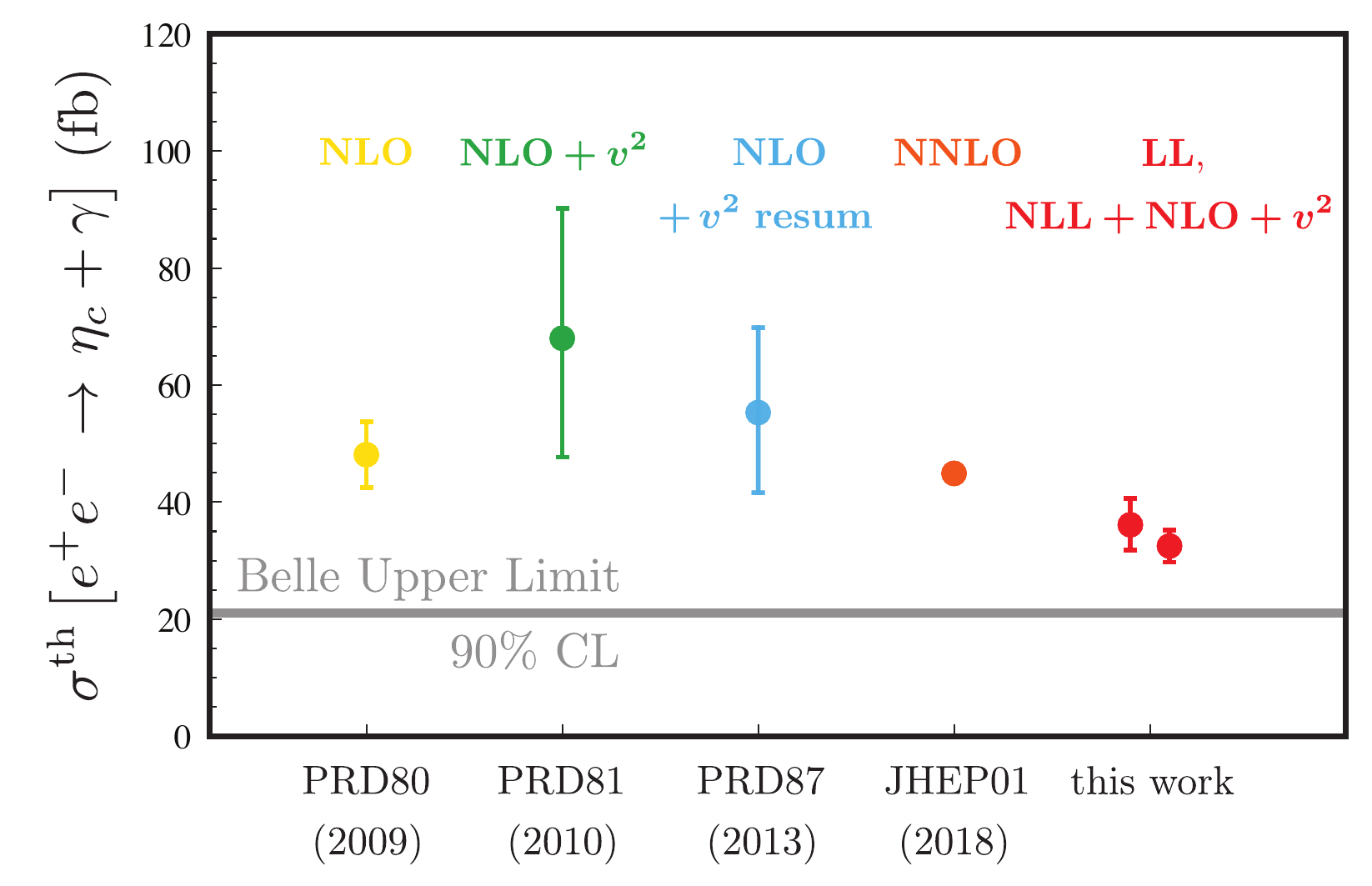}
\caption{Status of NRQCD predictions (points) and Belle's upper limit \cite{Jia:2018xsy} (horizontal gray line)  for $e^+e^- \to \eta_c+\gam$. This work on right side for LL and for NLL+NLO with $v^2$ corrections is compared to previous predictions (from the left) at NLO\cite{Li:2009ki}, NLO with $v^2$ correction \cite{Sang:2009jc} and with $v^2$ resummation\cite{Fan:2012dy}, and NNLO\cite{Chen:2017pyi}. We note that uncertainties of NLO\cite{Li:2009ki} and of NNLO\cite{Chen:2017pyi} include only quark mass variations while the others include other sources of uncertainties. See the text for more details. 
\label{fig:predictions}}
\end{figure}
In \fig{fig:predictions}, we summarize the status of NRQCD predictions (points) in comparison with Belle's upper limit (90\% credibility level) \cite{Jia:2018xsy} (gray line) for $\sigma(e^+e^-\to \eta_c+\gamma)$ at $\sqrt{s}=10.58$~GeV. Since the resummation effect is not substantial at this energy our results LL and NLL+NLO+$v^2$ on the right side of the plot should be comparable to LO and to NLO with $v^2$ corrections.

For a fair comparison with previous predictions we need to point out several major differences of input parameters and their variations between different predictions. First, a small error bar of NLO \cite{Li:2009ki} and invisibly small error of NNLO\cite{Chen:2017pyi} only include a charm-quark mass variation by $0.1$~GeV and they should not be compared to full uncertainties of other predictions. Instead their central values can be compared with the others.
Second, LO, NLO$+v^2$\cite{Sang:2009jc}, and NLO$+v^2$ resummation\cite{Fan:2012dy} use the LDME of \cite{Bodwin:2007fz}, which should be updated with improved measurement of $\Gamma[\eta_c\to \gam\gam]$ as discussed around \eq{eq:MEetac} and with the updated LDME, we expect decrease of the cross section by about $10\sim 20\%$ and also reduction of their uncertainties, quantitative estimation of which requires more careful study and is beyond scope of this paper. 
On the other hand our results of LL, NLL+NLO+$v^2$ in \fig{fig:predictions} is lower in its value and smaller in uncertainty partially due to this update.

There are differences in scale choice and its variation.
We use two scales $\mu=\sqrt{s}$ for the hard-scattering kernel and $\mu_0=2\msm_c$ for the LCDA and decay constant and they are varied by a factor of 2 for $\mu$ and $\sqrt{2}$ for $\mu_0$ as discussed in previous section.  While we use $\MSbar$ mass $\msm_c=1.275^{+0.25}_{-0.35}$, many of previous results use the pole mass $m_c=1.4\!\sim\!1.6$~GeV.
NLO \cite{Li:2009ki} sets $\mu=2 m_c$  
and NLO$+v^2$ resummation \cite{Fan:2012dy} sets $\mu=m_c, 2m_c$ (result with $2m_c$ is shown in \fig{fig:predictions} and the value for $m_c$ is similar), while NLO+$v^2$ \cite{Sang:2009jc} makes most conservative choice $\mu=\sqrt{s}, 2m_c,$ and $m_c$, which leads to relatively larger uncertainty compared that of NLO$+v^2$ resummation.
NNLO \cite{Chen:2017pyi} chooses different values for renormalization scale $\mu_r=\sqrt{s}/2$ and the factorization scale  $\mu_\Lambda= 1.0$~ GeV. We do not include the result of \cite{Braguta:2010mf} because the 1-loop coefficient is not consistent with other results \cite{Sang:2009jc,Wang:2013ywc}.

Recently the Belle experiment analyzed S-wave ($\eta_c+\gam$) and P-wave ($\chi_{cJ}+\gam$ with $J=0,1,2$) channels \cite{Jia:2018xsy}. While P-wave cross section ($J=1$) and upper limits ($J=0,2$) are consistent with the theoretical predictions \cite{Chung:2008km,Sang:2009jc,Li:2009ki,Braguta:2010mf}, S-wave upper limit $\sigma^\text{exp}_{\eta_c+\gam}\le 21.1$~fb at 90~\% credibility level is in tension with our NLL+NLO prediction $32.7\pm{2.8}$~fb by $4.1\sigma$.
This reminds us the puzzle in exclusive $J/\psi+\eta_c$ production \cite{Abe:2004ww,Aubert:2005tj,Braaten:2002fi, Liu:2002wq}, where a large discrepancy between theory and experiment was resolved by the combined effect of large $K$-factor, resummed relativistic corrections and careful determination of NRQCD matrix element \cite{Zhang:2005cha,Bodwin:2007fz,Bodwin:2007ga,He:2007te}.\footnote{Recently, \cite{Feng:2019zmt} reports the $K$-factor (NNLO/LO) between +20$\%$ and -40$\%$ depending on scale choice.} However, in our case effect of the $K$-factor, a ratio of NLL+NLO including relativistic correction relative to LO, is less than 5$\%$ as shown in \fig{fig:charm-2mc}. It would be surprising if higher-order resummation or relativistic corrections is the resolution to this tension. Of course, more careful study on those corrections and other contributions from different topology can shed lights on the tension. Without correct understanding of this channel one may also cast a doubt on theoretical prediction for other exclusive processes such as radiative Higgs decay into quarkonium, a novel channel to probe the Yukawa coupling of charm quark \cite{Sirunyan:2018fmm,Aaboud:2018txb,Bodwin:2013gca,Mao:2019hgg}. In this aspect, resolving the tension would be one of important checkpoints. The Belle II experiment with upgraded luminosity is starting its physics program and in a few years it will release improved measurements and can clarify if this seemingly tension is to be or not.

\section{Summary} \label{sec:summary}
We resum large logarithms of $4m_Q^2/s$ at NLL accuracy for exclusive production for $\eta_{b,c}+\gamma$ in high-energy lepton colliders by using light-cone factorization theorem and by using 2-loop evolution kernel  known from the pion form factor. The leading relativistic correction is also included and logarithms in the correction is resummed at LL accuracy.
 The nonsingular part of order $\as$ is obtained by subtracting the singular part from fixed-order results at NLO then, is added to resummed cross section. This makes our prediction of order $\as$ accuracy valid in both resummation region ($r \ll 1$) and fixed-order region ($r \sim O(1)$) where $r=4m_Q^2/s$ such that the results with the same formalism in \eq{eq:sigma} can be compared to measurement at the Belle energy near 10~$\GeV$ and in future colliders such as ILC, CEPC, FCC-ee.
 
Our final state $\eta_{c,b}$ is the pseudoscalar, which involves an ambiguity in handling $\gamma_5$ in $d$ dimension and the scheme dependency enters in individual parts such as hard kernel, LCDA, and the decay constant in factorized formula.
We explicitly showed that how the $\gamma_5$-scheme dependence vanishes in the resummed expression at NLL accuracy.
In resummed expression, there is a part proportional to fixed-order singular result and its scheme independence is followed by that of the fixed-order cross section.
In the other part of resummed expression, we observe that the scheme dependence of 2-loop anomalous dimension is matched to and cancelled against constant term of 1-loop hard-scattering kernel \eq{eq:Deltasum}.
 
In numerical calculation in \sec{results} we first rewrite the decay constant in terms of the experimental decay rate by eliminating NRQCD matrix element to avoid a systematic bias by unnecessary higher-order $\as$ contribution that can be contained in NRQCD matrix element. By doing this all the input formula are computed at the same $\as$ order to working accuracy and it is observed to show better perturbative convergence from LO to NLO and from LL to NLL. Our predictions for the cross sections and branching fraction are summarized in \tab{tab:sigmas}. The input decay rates $\Gamma[\eta_{b,c} \to \gamma \gamma]$ dominates over the others including perturbative uncertainty and uncertainty of our prediction reduces if the measurement of decay rate improves. In \sec{comparison} we compare our prediction to previous predictions for the Belle experiment and discussed the differences in input parameters and uncertainty estimates.  
We also find Belle's recent upper limit $21$~fb is about $4\,\sigma$ away from our prediction $33\pm 3$~fb.
We hope future Belle II analysis with better statistics coming-out in a few years and careful theoretical investigation on higher-order corrections may shed lights on this tension.

\begin{acknowledgments}
We thank Geoffrey T. Bodwin for sharing his knowledge on Abel-Pad\'e method
which was essential for resumming the non-convergent series appearing in 
the evolution of the heavy quarkonium LCDAs.
We also thank Chaehyun Yu and Yu Jia for useful conversations during the completion of this work.
The work of H.S.C.\ is supported by the Alexander von Humboldt Foundation.
The work of D.K.\ is supported by NSFC through Grant No.~11875112 and the work of X.W.\ is supported by Fudan Scholar program. 
The work of J.-H.E.\,\!, U-R.K.\,\!, and J.L.\ is supported by the National Research Foundation of Korea (NRF) under 
Contract No. NRF-2017R1E1A1A01074699 (J.-H.E.\,\!, U-R.K.\,\!, J.L.),
NRF-2018R1A2A3075605 (J.-H.E.\,\!, U-R.K.), NRF-2018R1D1A1B07047812 (U-R.K.),
NRF-2019R1A6A3A01096460 (U-R.K.), and NRF-2017R1A2B4011946 (J.-H.E.).
D.K.\ would like to thank the hospitality of QCD Group, Korea University where an important part of this work was carried out.
\end{acknowledgments}

\newpage
\appendix 
\section{fixed-order cross section}\label{app:fixedsig}

The fixed-order cross section up to $O(\as, v^2)$ was computed in~\cite{Sang:2009jc}
\be \label{eq:sigfixed} 
\sigma^\text{fixed}(r;\mu) = \sigma_0 \left[ (1-r)+\frac{\alpha_s C_F}{4\pi} c^\text{fixed} +\langle v^2\rangle c^\text{fixed}_{v^2}  \right] 
\,, \ee
where the coefficients  $c^\text{fixed}$ and $c^\text{fixed}_{v^2}$ are
\bea
c^\text{fixed}
&=&
-
\frac{2\left[30 r^2-(84+\pi^2)r+2\pi^2+54\right]}{3(2-r)}
+
\frac{8(2r-3)(1-r)}{(2-r)^2}
\log\left(\frac{2}{r}-2\right)
\nonumber\\&&
-\frac{12(1-r)}{\sqrt{1-r}}
\log\left(\frac{1-\sqrt{1-r}}{1+\sqrt{1-r}}\right)
-2
\left[
\left( 1 + \frac{r}{2} \right)\log^2\left(\frac{1-\sqrt{1-r}}{1+\sqrt{1-r}}\right)
-\log^2\left(\frac{2}{r}-1\right)
\right]
\nonumber\\&&
+4
\text{Li}_2\left(\frac{r}{2-r}\right)
\,,\nn\\
c^\text{fixed}_{v^2}&=&-\frac43\left(1-\frac r 4\right)
\,.\eea
Note that our coefficient $c^{\text{fixed}}$ is related to that of $C(r)$ in \cite{Sang:2009jc} as: $c^\text{fixed}=3(1-r)C(r)$.

\section{plus distributions} \label{app:plus}

Here, we give the definition of plus distributions used in the paper.
The $+$ and $++$ functions are defined by
\begin{eqnarray}
\label{eq:++-+-functions}
\int_0^1dx\,[f(x)]_+g(x)&=&\int_0^1 dx\,f(x)[g(x)-g(1/2)],
\nonumber \\
\int_0^1dx\,[f(x)]_{++}g(x)&=&\int_0^1 dx\,f(x)[g(x)-g(1/2)-g'(1/2)(x-1/2)],
\end{eqnarray}
The plus distribution depending on two arguments $x$ and $y$ is defined by
\begin{equation}
\left[f(x,y)\right]_+
\equiv
f(x,y)-\delta(x-y)\int_0^1 dz f(z,y).
\end{equation}

\section{anomalous dimension \label{app:anomalous-dim}}

The NLO anomalous dimension $\gamma_{n-1}^{\parallel(1)}$ is given in 
Ref.~\cite{GonzalezArroyo:1979df} as
\begin{eqnarray}
\label{eq:two-loop-anomalous-dim}
\gamma_{n-1}^{(1)}
&=&
\left(C_F^2-\frac{1}{2}C_F C_A\right)
\Bigg\{
16H_n\frac{2n+1}{n^2(n+1)^2}
+16\left[2H_n-\frac{1}{n(n+1)}
\right]
\left(H_n^{(2)}-S_{n/2}^{'(2)}\right)
\nonumber \\
&&\quad\quad\quad\quad\quad\quad\quad\quad
+64\tilde{S}_n+24H_n^{(2)}-3-8S_{n/2}^{'(3)}
-8\frac{3n^3+n^2-1}{n^3(n+1)^3}
-16(-1)^n \frac{2n^2+2n+1}{n^3(n+1)^3}
\Bigg\}
\nonumber \\
&&
+C_FC_A
\Bigg\{
H_n
\left[
\frac{536}{9}+8\frac{2n+1}{n^2(n+1)^2}
\right]
-16H_n H_n^{(2)}
+H_n^{(2)}
\left[
-\frac{52}{3}+\frac{8}{n(n+1)}
\right]
\nonumber \\
&&\quad\quad\quad\quad
-\frac{43}{6}
-4\frac{151n^4+263n^3+97n^2+3n+9}{9n^3(n+1)^3}
\Bigg\}
\nonumber \\
&&
+C_F\frac{n_f}{2}
\Bigg\{
-\frac{160}{9}H_n+\frac{32}{3}H_n^{(2)}
+\frac{4}{3}
+16\frac{11n^2+5n-3}{9n^2(n+1)^2}
\Bigg\},
\end{eqnarray}
where
\begin{eqnarray}
H_{n}^{(k)}&\equiv&\sum_{j=1}^n \frac{1}{j^k},
\quad
\textrm{with}
\quad
H_n^{(1)}\equiv H_{n},
\\
S^{'(k)}_{n/2}&\equiv&
\begin{cases}
\displaystyle
H^{(k)}_{n/2},
& \textrm{if $n$ is even,}
\\
\displaystyle
H^{(k)}_{(n-1)/2},
& \textrm{if $n$ is odd,}
\end{cases}
\\
\tilde{S}_n
&\equiv&
\sum_{j=1}^n \frac{(-1)^j}{j^2}H_j.
\end{eqnarray}
The off-diagonal evolution factor $d_{nk}(\mu,\mu_0)$ is given by
\begin{equation}
d_{nk}(\mu,\mu_0)
=
\frac{M_{nk}}
{\gamma_n^{(0)}-\gamma_k^{(0)}-2\beta_0}
\left\{
1-\left[
\frac{\alpha_s(\mu)}{\alpha_s(\mu_0)}
\right]^{\frac{\gamma_n^{(0)}-\gamma_k^{(0)}
-2\beta_0}{2\beta_0}}
\right\},
\end{equation}
where
\begin{eqnarray}
M_{nk}&=&
\frac{(k+1)(k+2)(k+3)}{(n+1)(n+2)}
(\gamma_n^{(0)}-\gamma_k^{(0)})
\left[
\frac{8C_F A_{nk}-\gamma_k^{(0)}-2\beta_0}
{(n-k)(n+k+3)}
+4C_F\frac{A_{nk}-\psi(n+2)+\psi(1)}{(k+1)(k+2)}
\right],
\nonumber \\
A_{nk}&=&
\psi\left(\frac{n+k+4}{2}\right)-\psi\left(\frac{n-k}{2}\right)
+2\psi(n-k)-\psi(n+2)-\psi(1),
\end{eqnarray}
and $\psi(n)$ is the digamma function.

\newpage
\bibliography{paper}
\end{document}